# Resource Allocation Among Agents with MDP-Induced Preferences

**Dmitri A. Dolgov**　　　　　　　　　　　　　　　　DDOLGOV@AI.STANFORD.EDU
*Technical Research Department (AI & Robotics Group)*
*Toyota Technical Center*
*2350 Green Road*
*Ann Arbor, MI 48105, USA*

**Edmund H. Durfee**　　　　　　　　　　　　　　　　DURFEE@UMICH.EDU
*Electrical Engineering and Computer Science*
*University of Michigan*
*2260 Hayward St.*
*Ann Arbor, MI 48109, USA*

## Abstract

Allocating scarce resources among agents to maximize global utility is, in general, computationally challenging. We focus on problems where resources enable agents to execute actions in stochastic environments, modeled as Markov decision processes (MDPs), such that the value of a resource bundle is defined as the expected value of the optimal MDP policy realizable given these resources. We present an algorithm that simultaneously solves the resource-allocation and the policy-optimization problems. This allows us to avoid explicitly representing utilities over exponentially many resource bundles, leading to drastic (often exponential) reductions in computational complexity. We then use this algorithm in the context of self-interested agents to design a combinatorial auction for allocating resources. We empirically demonstrate the effectiveness of our approach by showing that it can, in minutes, optimally solve problems for which a straightforward combinatorial resource-allocation technique would require the agents to enumerate up to $2^{100}$ resource bundles and the auctioneer to solve an NP-complete problem with an input of that size.

## 1. Introduction

The problem of resource allocation is ubiquitous in many diverse research fields such as economics, operations research, and computer science, with applications ranging from decentralized scheduling (e.g., Wellman, Walsh, Wurman, & MacKie-Mason, 2001) and network routing (e.g., Feldmann, Gairing, Lucking, Monien, & Rode, 2003) to transportation logistics (e.g., Sheffi, 2004; Song & Regan, 2002) and bandwidth allocation (e.g., McMillan, 1994; McAfee & McMillan, 1996), just to name a few. The core question in resource allocation is how to distribute a set of scarce resources among a set of agents (either cooperative or self-interested) in a way that maximizes some measure of global utility, with social welfare (sum of agents' utilities) being one of the most popular criteria.

In many domains, an agent's utility for obtaining a set of resources is defined by what the agent can accomplish using these resources. For example, the value of a vehicle to a delivery agent is defined by the additional revenue that the agent can obtain by using the vehicle. However, to figure out how to best utilize a resource (or set of resources), an agent often must solve a non-trivial planning problem because actions might have long-term, non-deterministic effects. Therefore, an agent's value for a set of resources is defined by a solution to a planning problem, but to formulate its planning problem the agent needs to know what resources it will obtain. This leads to cyclic dependencies (depicted in Figure 1), wherein the input to the resource allocation problem depends on the solution to the planning problem, and vice versa. Unfortunately, for anything but the simplest domains, neither the resource-allocation nor the planning problem can be solved in closed form, making it impossible to obtain parameterized solutions.

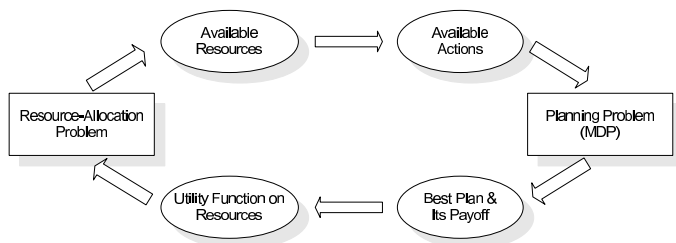

Figure 1: Dependency Cycle: To formulate their planning problems, the agents need to know what resources they will get, but their utility functions, which define the input to the resource-allocation problem, depend on the solutions to the planning problems.

Our focus in this paper is on solving the interdependent problems of resource allocation and stochastic planning. The main question we consider is how to allocate the resources in a way that maximizes the social welfare of the agents when the utility function of each agent is defined by a Markov decision process (Puterman, 1994) whose action set is parameterized by the resources. In this paper, we specifically focus on non-consumable resources (such as vehicles) that enable actions, but are not themselves consumed during action execution. We briefly mention the case of consumable resources in Section 6, and refer to the work by Dolgov (2006) for a more detailed treatment.

We assume that the agents' MDPs are weakly-coupled, meaning the agents only interact through the resources, and once the resources are allocated, the transition and reward functions of their MDPs are independent. Our model of weakly-coupled MDPs connected via shared resources is similar to that of Meuleau, Hauskrecht, Kim, Peshkin, Kaelbling, Dean, and Boutilier (1998) and Benazera, Brafman, Meuleau, and Hansen (2005), but differs in that we further assume that resources are only allocated once, prior to any actions being taken. While this "one-shot" allocation assumption limits our approach somewhat, it also allows our approach to apply more broadly to non-cooperative settings (without this assumption, the game-theoretic analysis of agents' interactions is significantly more complex). More importantly, it allows us to avoid the state space explosion (due to including resource information in the MDP states), which limits that other work to finding only approximately optimal solutions for non-trivial problems.

The main result presented in this paper is thus a new algorithm that, under the above conditions, *optimally solves the resource-allocation and the policy-optimization problems simultaneously*. By considering the two problems together, it sidesteps the dependency cycle mentioned above, which allows us to avoid an explicit representation of utility functions on resource bundles, leading to an exponential reduction in complexity over combinatorial resource allocation with flat utility functions. We empirically demonstrate that the resulting algorithm scales well to finding optimal solutions for problems involving numerous agents and resources.

Our algorithm can be viewed as contributing a new approach for dealing with the computational complexity of resource allocation in domains with complex utility functions that are not linearly decomposable in the resources (due to the effects of substitutability and complementarity). In such *combinatorial allocation problems*, finding an optimal allocation is NP-complete in the (often exponentially large) space of resource bundles (Rothkopf, Pekec, & Harstad, 1998). Previous approaches for addressing the complexity have included determining classes of utility functions that lead to tractable problems (as surveyed by de Vries & Vohra, 2003), iterative algorithms for resource allocation and preference elicitation (as surveyed by Sandholm & Boutilier, 2006), and concise languages for expressing agents' preferences (Sandholm, 1999; Nisan, 2000; Boutilier & Hoos, 2001; Boutilier, 2002). The novelty of our approach with respect to these is that it explicitly embraces the underlying processes that define the agents' utility functions, for cases where these processes can be modeled as resource-parameterized MDPs. By doing so, not only does our approach use such MDP-based models as a concise language for agents' utility functions, but more importantly, it directly exploits the structure in these models to drastically reduce computational complexity by simultaneously solving the planning and resource-allocation problems.

In the context of cooperative agents, our approach can be viewed as a way of solving weakly-coupled multiagent MDPs, where agents' transition and reward functions are independent, but the space of joint actions is constrained, as, for example, in the models used by Singh and Cohn (1998) or Meuleau et al. (1998). From that perspective, the concept of resources can be viewed as a compact way of representing the interactions between agents, similarly to the model used by Bererton, Gordon, and Thrun (2003); however, our work differs in a number of assumptions. Moreover, our algorithms can be easily modified to work with models where the constraints on the joint actions are modeled directly (for example, via SAT formulas).

For non-cooperative agents, we apply our resource-allocation algorithm to the mechanism-design problem (e.g., Mas-Colell, Whinston, & Green, 1995), where the goal is to allocate resources among the agents in a way that maximizes the social welfare, given that each participating agent is selfishly maximizing its own utility. For domains with self-interested agents with complex preferences that exhibit combinatorial effects between the resources, *combinatorial auctions* (e.g., de Vries & Vohra, 2003) are often used for resource-allocation. The Generalized Vickrey Auction (GVA) (MacKie-Mason & Varian, 1994), which is an extension of Vickrey-Clarke-Groves (VCG) mechanisms (Vickrey, 1961; Clarke, 1971; Groves, 1973) to combinatorial auctions, is particularly attractive because of its nice analytical properties (as described in Section 4.1). We develop a variant of a VCG auction, where agents submit their resource-parameterized MDPs as bids, and the auctioneer simultaneously solves the resource-allocation and the policy-optimization problems, thus retaining the compact representation of agents' preferences throughout the process. We describe extensions to the mechanism for distributing the computation and for encoding MDP information to reduce the revelation of private information.

The remainder of this paper proceeds as follows. After a brief review of MDPs in Section 2, we present (in Section 3) our model of a decision-making agent: the resource-parameterized MDP with capacity constraints. We analyze the problem of optimal policy formulation in such a resource-parameterized capacity-constrained MDP, study its properties, and present a solution algorithm, based on the formulation of this (NP-complete) problem as a mixed integer program.

With these building blocks, we move to the multiagent setting and present our main result, the algorithm for simultaneously allocating resources and planning across agents (Section 4). Based on that algorithm, we then design a combinatorial auction for allocating resources among self-interested agents. We describe a distributed implementation of the mechanism, and discuss techniques for preserving information privacy. In Section 5, we analyze the computational efficiency of our approach, empirically demonstrating exponential reductions in computational complexity, compared to a straightforward combinatorial resource-allocation algorithm with flat utility functions. Finally, in Section 6, we conclude with a discussion of possible generalizations and extensions to our approach. For conciseness and better readability, all proofs and some generalizations are deferred to appendices.

## 2. Markov Decision Processes

We base our model of agents' decision problems on infinite-horizon fully-observable MDPs with the total expected discounted reward optimization criterion (although our results are also applicable to other classes of MDPs, such as MDPs with the average per-step rewards). This section introduces our notation and assumptions, and serves as a brief overview of the basic MDP results (see, for example, the text by Puterman (1994) for a detailed discussion of the material in this section).

A classical single-agent, unconstrained, stationary, fully-observable MDP can be defined as a 4-tuple $\langle \mathcal{S}, \mathcal{A}, p, r \rangle$, where:

- $\mathcal{S}$ is a finite set of states the agent can be in.

- $\mathcal{A}$ is a finite set of actions the agent can execute.

- $p : \mathcal{S} \times \mathcal{A} \times \mathcal{S} \mapsto [0, 1]$ defines the transition function. The probability that the agent goes to state $\sigma \in \mathcal{S}$ upon execution of action $a \in \mathcal{A}$ in state $s \in \mathcal{S}$ is $p(\sigma | s, a)$. We assume that, for any action, the corresponding transition matrix is stochastic: $\sum_\sigma p(\sigma | s, a) = 1 \ \forall s \in \mathcal{S}, a \in \mathcal{A}$.

- $r : \mathcal{S} \times \mathcal{A} \mapsto \mathbb{R}$ defines the reward function. The agent obtains a reward of $r(s, a)$ if it executes action $a \in \mathcal{A}$ in state $s \in \mathcal{S}$. We assume the rewards are bounded.

In a discrete-time fully-observable MDP, at each time step, the agent observes the current state of the system and chooses an action according to its *policy*. A policy is said to be *Markovian* (or *history-independent*) if the choice of action does not depend on the history of states and actions encountered in the past, but rather only on the current state and time. If, in addition to that, the policy does not depend on time, it is called *stationary*. By definition, a stationary policy is always Markovian. A *deterministic* policy always prescribes the execution of the same action in a state, while a *randomized* policy chooses actions according to a probability distribution.

Following the standard notation (Puterman, 1994), we refer to different classes of policies as $\Pi^{xy}$, where $x = \{H, M, S\}$ specifies whether a policy is History-dependent, Markovian, or Stationary, and $y = \{R, D\}$ specifies whether the policy is Randomized or Deterministic (e.g., the class of stationary deterministic policies is labeled $\Pi^{SD}$). Obviously, $\Pi^{Hy} \supset \Pi^{My} \supset \Pi^{Sy}$ and $\Pi^{xR} \supset \Pi^{xD}$, with history-dependent randomized policies $\Pi^{HR}$ and stationary deterministic policies $\Pi^{SD}$ being the most and the least general, respectively.

A stationary randomized policy is thus a mapping of states to probability distributions over actions: $\pi : \mathcal{S} \times \mathcal{A} \mapsto [0, 1]$, where $\pi(s, a)$ defines the probability that action $a$ is executed in state $s$. A stationary deterministic policy can be viewed as a degenerate case of a randomized policy for which there is only one action for each state that has a nonzero probability of being executed.

In an unconstrained discounted MDP, the goal is to find a policy that maximizes the total expected discounted reward over an infinite time horizon:[1]

$$U_\gamma(\pi, \alpha) = \mathbb{E}\Big[ \sum_{t=0}^{\infty} (\gamma)^t r_t(\pi, \alpha) \Big], \tag{1}$$

where $\gamma \in [0, 1)$ is the discount factor (a unit reward at time $t + 1$ is worth the same to the agent as a reward of $\gamma$ at time $t$), $r_t$ is the (random) reward the agent receives at time $t$, whose distribution depends on the policy $\pi$ and the initial distribution over the state space $\alpha : \mathcal{S} \mapsto [0, 1]$.

One of the most important results in the theory of MDPs states that, for an unconstrained discounted MDP with the total expected reward optimization criterion, there always exists an optimal policy that is stationary, deterministic, and *uniformly optimal*, where the latter term means that the policy is optimal for all distributions over the starting state.[2]

There are several commonly-used ways of finding the optimal policy, and central to all of them is the concept of a *value function* of a policy, $v_\pi : \mathcal{S} \mapsto \mathbb{R}$, where $v_\pi(s)$ is the expected cumulative value of the reward the agent would receive if it started in state $s$ and behaved according to policy $\pi$. For a given policy $\pi$, the value of every state is the unique solution to the following system of $|\mathcal{S}|$ linear equations:

$$v_\pi(s) = \sum_a r(s, a)\pi(s, a) + \gamma \sum_\sigma p(\sigma|s, a)v_\pi(\sigma), \qquad \forall s \in \mathcal{S}. \tag{2}$$

To find the optimal policy, it is handy to consider the *optimal value function* $v^* : \mathcal{S} \mapsto \mathbb{R}$, where $v^*(s)$ represents the value of state $s$, given that the agent behaves optimally. The optimal value function satisfies the following system of $|\mathcal{S}|$ nonlinear equations:

$$v^*(s) = \max_a \Big[ r(s, a) + \gamma \sum_\sigma p(\sigma|s, a)v^*(\sigma) \Big], \qquad \forall s \in \mathcal{S}. \tag{3}$$

Given the optimal value function $v^*$, an optimal policy is to simply act greedily with respect to $v^*$ (with any method of tie-breaking in case of multiple optimal actions):

$$\pi(s, a) = \begin{cases} 1 & \text{if } a \in \arg\max_a \Big[ r(s, a) + \gamma \sum_\sigma p(\sigma|s, a)v^*(\sigma) \Big], \\ 0 & \text{otherwise.} \end{cases} \tag{4}$$

---

1. Notation: here and below $(x)^y$ is an exponent, while $x^y$ is a superscript.
2. Uniform optimality of policies is the reason why $\alpha$ is not included as a component of a textbook MDP.

One of the possible ways of solving for the optimal value function is to formulate the nonlinear system (3) as a linear program (LP) with $|\mathcal{S}|$ optimization variables $v(s)$ and $|\mathcal{S}||\mathcal{A}|$ constraints:

$$\min \sum_s \alpha(s)v(s)$$

subject to:

$$v(s) \geq r(s,a) + \gamma \sum_\sigma p(\sigma|s,a)v(\sigma), \qquad \forall s \in \mathcal{S}, a \in \mathcal{A},$$

(5)

where $\alpha$ is an arbitrary constant vector with $|\mathcal{S}|$ positive components ($\alpha(s) > 0 \ \forall s \in \mathcal{S}$).[3]

In many problems (including the ones that are the focus of this paper), it is very useful to consider the equivalent dual LP with $|\mathcal{S}||\mathcal{A}|$ optimization variables $x(s,a)$ and $|\mathcal{S}|$ constraints:[4]

$$\max_x \sum_s \sum_a r(s,a)x(s,a)$$

subject to:

$$\sum_a x(\sigma,a) - \gamma \sum_s \sum_a x(s,a)p(\sigma|s,a) = \alpha(\sigma), \qquad \forall \sigma \in \mathcal{S};$$

$$x(s,a) \geq 0 \qquad\qquad\qquad\qquad \forall s \in \mathcal{S}, a \in \mathcal{A}.$$

(6)

The optimization variables $x(s,a)$ are often called the *visitation frequencies* or the *occupation measure* of a policy. If we think of $\alpha$ as the initial probability distribution, then $x(s,a)$ can be interpreted as the total expected number of times action $a$ is executed in state $s$. Then, $x(s) = \sum_a x(s,a)$ gives the total expected *flow* through state $s$, and the constraints in the above LP can be interpreted as the conservation of flow through each of the states.

An optimal policy can be computed from a solution to the dual LP as:

$$\pi(s,a) = \frac{x(s,a)}{\sum_a x(s,a)},$$

(7)

where non-negativity of $\alpha$ guarantees that $\sum_a x(s,a) > 0 \ \forall s \in \mathcal{S}$. In general, this appears to lead to randomized policies. However, a bounded LP with $n$ constraints always has a *basic feasible solution* (e.g., Bertsimas & Tsitsiklis, 1997), which by definition has no more than $n$ non-zero components. If $\alpha$ is strictly positive, a basic feasible solution to the LP (7) will have precisely $|S|$ nonzero components (one for each state), which guarantees the existence of an optimal deterministic policy. Such a policy can be easily obtained by most LP solvers (e.g., simplex will always produce solutions that map to deterministic policies).

Furthermore, as mentioned above, for unconstrained discounted MDPs, there always exist policies that are uniformly optimal (optimal for all initial distributions). The above dual LP (6) yields uniformly optimal policies if a strictly positive $\alpha$ is used. However, the solution $(x)$ to the dual LP retains its interpretation as the expected number of times state $s$ is visited and action $a$ is executed only for the initial probability distribution $\alpha$ that was used in the LP.

---

3. The overloading of $\alpha$ as the objective function coefficients here and the initial probability distribution of an MDP earlier is intentional and is explained shortly.

4. Note that some authors (e.g., Altman, 1996) prefer the opposite convention, where (6) is called the dual, and (5), the primal.

The main benefit of the dual LP (6) is manifested in constrained MDPs (Altman, 1999; Kallenberg, 1983; Heyman & Sobel, 1984), where each action, in addition to producing a reward $r(s, a)$, also incurs a vector of costs $\eta_k(s, a) : \mathcal{S} \times \mathcal{A} \mapsto \mathbb{R} \; \forall k \in [1..K]$. The problem then is to maximize the expected reward, subject to constraints on the expected costs. Constrained models of this type arise in many domains, such as telecommunication applications (e.g., Ross & Chen, 1988; Ross & Varadarajan, 1989), where it is often desirable to maximize expected throughput, subject to conditions on the average delay. Such problems, where constraints are imposed on the expected costs, can be solved in polynomial time using linear programming by simply augmenting the dual LP (6) with the following linear constraints:

$$\sum_s \sum_a \eta_k(s, a) x(s, a) \leq \widehat{\eta}_k, \qquad \forall k \in [1..K], \tag{8}$$

where $\widehat{\eta}_k$ is the upper bound on the expected cost of type $k$. The resulting constrained MDP differs from the standard unconstrained MDP: in particular, deterministic policies are no longer optimal, and uniformly optimal policies do not, in general, exist for such problems (Kallenberg, 1983).

For the same reason of being easily augmentable with constraints, the dual LP (6) also forms the basis for our approach. However, the constraints that arise in resource-allocation problems that are our focus in this paper are very different from the linear constraints in (8), leading to different optimization problems with different properties and requiring different solution techniques (as described in more detail in Section 3).

We conclude the background section by introducing a simple unconstrained MDP that will serve as the basis for a running example, to which we will refer throughout the rest of this paper.

**Example 1** *Consider a simple delivery domain, depicted in Figure 2, where the agent can obtain rewards for delivering furniture (action $a_1$) or delivering appliances (action $a_2$). Delivering appliances produces higher rewards (as shown on the diagram), but it does more damage to the delivery vehicle. If the agent only delivers furniture, the damage to the vehicle is negligible, whereas if the agent delivers appliances, the vehicle is guaranteed to function reliably for the first year (state $s_1$), but after that (state $s_2$) has a 10% probability of failure, per year. The vehicle can be serviced (action $a_3$), resetting its condition, but at the expense of lowering profits. If the truck does break (state $s_3$), it can be repaired (action $a_4$), but with a more significant negative impact on profits. We will assume a discount factor $\gamma = 0.9$.*

*The optimal value function is $v^*(s_1) \approx 95.3, v^*(s_2) \approx 94.7, v^*(s_3) \approx 86.7$, and the corresponding optimal occupation measure (assuming a uniform $\alpha$) is the following (listing only the non-zero elements): $x(s_1, a_2) \approx 4.9, x(s_2, a_3) \approx 4.8, x(s_3, a_4) \approx 0.3$. This maps to the optimal policy that dictates that the agent is to start by delivering appliances (action $a_2$ in state $s_1$), then service the vehicle after the first year (action $a_3$ in state $s_2$), and fix the vehicle if it ever gets broken ($a_4$ in $s_3$) (the latter has zero probability of happening under this policy if the agent starts in state $s_1$ or $s_2$).* □

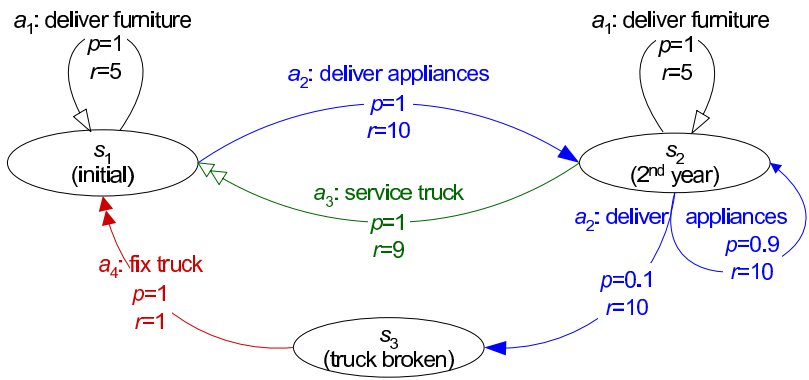

Figure 2: Unconstrained MDP example for a delivery domain. Transition probabilities ($p$) and rewards ($r$) are shown on the diagram. Actions not shown result in transition to same state with no reward. There is also a noop action $a_0$ that corresponds to doing nothing; it does not change state and produces zero reward.

## 3. Agent Model: Resource-Parameterized MDP

In this section, we introduce our model of the decision-making agent, and describe the single-agent stochastic policy-optimization problem that defines the agent's preferences over resources. We show that, for this single-agent problem, formulated as an MDP whose action set is parameterized on the resources available to the agent, stationary deterministic policies are optimal, but uniformly optimal policies do not, in general, exist. We also show that the problem of finding optimal policies is NP-complete. Finally, we present a policy-optimization algorithm, based on a formulation of the problem as a mixed integer linear program (MILP).

We model the agent's resource-parameterized MDP as follows. The agent has a set of actions that are potentially executable, and each action requires a certain combination of resources. To capture local constraints on the sets of resources an agent can use, we use the concept of capacities: each resource has capacity costs associated with it, and each agent has capacity constraints. For example, a delivery company needs vehicles and loading equipment (resources to be allocated) to make its deliveries (execute actions). However, all equipment costs money and requires manpower to operate it (the agent's local capacity costs). Therefore, the amount of equipment the agent can acquire and successfully utilize is constrained by factors such as its budget and limited manpower (agent's local capacity bounds). This two-layer model with capacities and resources represented separately might seem unnecessarily complex (why not fold them together or impose constraints directly on resources?), but the separation becomes evident and useful in the multiagent model discussed in Section 4. We emphasize the difference here: resources are the items being allocated among the agents, while capacities define the inherent limitations of an individual agent on what combinations of resources it can usefully possess.

The agent's optimization problem is to choose a subset of the available resources that does not violate its capacity constraints, such that the best policy feasible under that bundle of resources yields the highest utility. In other words, the single-agent problem analyzed in this section has no constraints on the total resource amounts (they are introduced in the multiagent problem in the next section), and the constraints are only due to the agent's capacity limits. Adding limited resource amounts to the single-agent model would be a very simple matter, since such constraints can be handled with a simple pruning of the agent's action space. Further, note that without capacity constraints, the single-agent problem would be trivial, as it would always be optimal for the agent to simply take all resources that are of potential use. However, in the presence of capacity constraints, we face a problem that is similar to the cyclic dependency in Figure 1: the resource-selection problem requires knowing the values of all resource bundles, which are defined by the planning problem, and the planning problem is ill-defined until a resource bundle is chosen. In this section, we focus on the single-agent problem of selecting an optimal subset of resources that satisfies the agent's capacity constraints and assume the agent has no value for acquiring additional resources that exceed its capacity bounds.

The resources in the model outlined above are non-consumable, i.e., actions require resources but do not consume them during execution. As mentioned in the Introduction, in this work we focus only on non-consumable resources, and only briefly outline the case of consumable resources in Section 6.

We can model the agent's optimization problem as an $n$-tuple $\langle \mathcal{S}, \mathcal{A}, p, r, \mathcal{O}, \rho, \mathcal{C}, \kappa, \widehat{\kappa}, \alpha \rangle$:

- $\langle \mathcal{S}, \mathcal{A}, p, r \rangle$ are the standard components of an MDP, as defined earlier in Section 2.

- $\mathcal{O}$ is the set of resources (e.g., $\mathcal{O} = \{$production equipment, vehicle, . . .$\}$). We will use $o \in \mathcal{O}$ to refer to a resource type.

- $\rho : \mathcal{A} \times \mathcal{O} \mapsto \mathbb{R}$ is a function that specifies the resource requirements of all actions; $\rho(a, o)$ defines how much of resource $o \in \mathcal{O}$ action $a \in \mathcal{A}$ needs to be executable (e.g., $\rho(a, \text{vehicle}) = 1$ means that action $a$ requires one vehicle).

- $\mathcal{C}$ is the set of capacities of our agent (e.g., $\mathcal{C} = \{$space, money, manpower, . . .$\}$). We will use $c \in \mathcal{C}$ to refer to a capacity type.

- $\kappa : \mathcal{O} \times \mathcal{C} \mapsto \mathbb{R}$ is a function that specifies the capacity costs of resources; $\kappa(o, c)$ defines how much of capacity $c \in \mathcal{C}$ a unit of resource $o \in \mathcal{O}$ requires (e.g., $\kappa(\text{vehicle}, \text{money}) = \$50000$ defines the monetary cost of a vehicle, while $\kappa(\text{vehicle}, \text{manpower}) = 2$ means that two people are required to operate the vehicle).

- $\widehat{\kappa} : \mathcal{C} \mapsto \mathbb{R}$ specifies the upper bound on the capacities; $\widehat{\kappa}(c)$ gives the upper bound on capacity $c \in \mathcal{C}$ (e.g., $\widehat{\kappa}(\text{money}) = \$1,000,000$ defines the budget constraint, and $\widehat{\kappa}(\text{manpower}) = 7$ specifies the size of the workforce).

- $\alpha : \mathcal{S} \mapsto \mathbb{R}$ is the initial probability distribution; $\alpha(s)$ is the probability that the agent starts in state $s$.

Our goal is to find a policy $\pi$ that yields the highest expected reward, under the conditions that the resource requirements of that policy do not exceed the capacity bounds of the agent. In other words, we have to solve the following mathematical program:[5]

$$\max_{\pi} U_\gamma(\pi, \alpha)$$

subject to:

$$\sum_o \kappa(o, c) \max_a \left\{ \rho(a, o) H\left(\sum_s \pi(s, a)\right) \right\} \leq \widehat{\kappa}(c), \qquad \forall c \in \mathcal{C},$$

$$(9)$$

where $H$ is the Heaviside "step" function of a nonnegative argument, defined as:

$$H(z) = \begin{cases} 0 & z = 0, \\ 1 & z > 0. \end{cases}$$

The constraint in (9) can be interpreted as follows. The argument of $H$ is nonzero if the policy $\pi$ assigns a nonzero probability to using action $a$ in at least one state. Thus, $H(\sum_s \pi(s, a))$ serves as an indicator function that tells us whether the agent plans to use action $a$ in its policy, and $\max \left\{ \rho(a, o) H(\sum_s \pi(s, a)) \right\}$ tells us how much of resource $o$ the agent needs for its policy. We take a max with respect to $a$, because the same resource $o$ can be used by different actions. Therefore, when summed over all resources $o$, the left-hand side gives us the total requirements of policy $\pi$ in terms of capacity $c$, which has to be no greater than the bound $\widehat{\kappa}(c)$.

The following example illustrates the single-agent model.

**Example 2** *Let us augment Example 1 as follows. Suppose the agents needs to obtain a truck to perform its delivery actions ($a_1$ and $a_2$). The truck is also required by the service and repair actions ($a_3$ and $a_4$). Further, to deliver appliances, the agent needs to acquire a forklift, and it needs to hire a mechanic to be able to repair the vehicle ($a_4$). The noop action $a_0$ requires no resources. This maps to a model with three resources (truck, forklift, and mechanic): $\mathcal{O} = \{o_t, o_f, o_m\}$, and the following action resource costs (listing only the non-zero ones):*

$$\rho(a_1, o_t) = 1, \ \rho(a_2, o_t) = 1, \ \rho(a_2, o_f) = 1, \ \rho(a_3, o_t) = 1, \ \rho(a_4, o_t) = 1, \ \rho(a_4, o_m) = 1.$$

*Moreover, suppose that the resources (truck $o_t$, forklift $o_f$, or mechanic $o_m$) have the following capacity costs (there is only one capacity type, money: $\mathcal{C} = \{c_1\}$)*

$$\kappa(o_t, c_1) = 2, \ \kappa(o_f, c_1) = 3, \ \kappa(o_m, c_1) = 4,$$

*and the agent has a limited budget of $\widehat{\kappa} = 8$. It can, therefore acquire no more than two of the three resources, which means that the optimal solution to the unconstrained problem as in Example 1 is no longer feasible.* □

Let us observe that the MDP-based model of agents' preferences presented above is fully general for discrete indivisible resources, i.e., any non-decreasing utility function over resource bundles can be represented via the resource-constrained MDP model described above.

---

5. This formulation assumes a stationary policy, which is supported by the argument in Section 3.1.

**Theorem 1** *Consider a finite set of $n$ indivisible resources $\mathcal{O} = \{o_i\}$ ($i \in [1, n]$), with $m \in \mathbb{N}$ available units of each resource. Then, for any non-decreasing utility function defined over resource bundles $f : [0, m]^n \mapsto \mathbb{R}$, there exists a resource-constrained MDP $\langle \mathcal{S}, \mathcal{A}, p, r, \mathcal{O}, \rho, \mathcal{C}, \kappa, \widehat{\kappa}, \alpha \rangle$ (with the same resource set $\mathcal{O}$) whose induced utility function over the resource bundles is the same as $f$. In other words, for every resource bundle $\mathbf{z} \in [0, m]^n$, the value of the optimal policy among those whose resource requirements do not exceed $\mathbf{z}$ (call this set $\Pi(\mathbf{z})$) is the same as $f(\mathbf{z})$.*

$$\forall f : [0, m]^n \mapsto \mathbb{R}, \; \exists \; \langle \mathcal{S}, \mathcal{A}, p, r, \mathcal{O}, \rho, \mathcal{C}, \kappa, \widehat{\kappa}, \alpha \rangle :$$

$$\forall \; \mathbf{z} \in [0, m]^n, \Pi(\mathbf{z}) = \left\{ \pi \, \Big| \, \max_a \left[ \rho(a, o_i) H\left( \sum_s \pi(s, a) \right) \right] \leq z_i \right\} \implies \max_{\pi \in \Pi(\mathbf{z})} U_\gamma(\pi, \alpha) = f(\mathbf{z}).$$

**Proof:** See Appendix A.1. □

Let us comment that while Theorem 1 establishes the generality of the MDP-based preference model introduced in this section, the construction used in the proof is of little practical interest, as it requires an MDP with an exponentially large state or action space. Indeed, we do not advocate mapping arbitrary unstructured utility functions to exponentially-large MDPs as a general solution technique. Rather, our contention is that our techniques apply to domains where utility functions are induced by a stochastic decision-making process (modeled as an MDP), thus resulting in well-structured preferences over resources that can be exploited to drastically lower the computational complexity of resource-allocation algorithms.

## 3.1 Properties of the Single-Agent Constrained MDP

In this section, we analyze the constrained policy-optimization problem (9). Namely, we show that stationary deterministic policies are optimal for this problem, meaning that it is not necessary to consider randomized, or history-dependent policies. However, solutions to problem (9) are not, in general, uniformly optimal (optimal for any initial distribution). Furthermore, we show that (9) is NP-hard, unlike the unconstrained MDPs, which can be solved in polynomial time (Littman, Dean, & Kaelbling, 1995).

We begin by showing optimality of stationary deterministic policies for (9). In the following, we use $\Pi^{\text{HR}}$ to refer to the class of history-dependent randomized policies (the most general policies), and $\Pi^{\text{SD}} \subset \Pi^{\text{HR}}$ to refer to the class of stationary deterministic policies.

**Theorem 2** *Given an MDP $M = \langle \mathcal{S}, \mathcal{A}, p, r, \mathcal{O}, \rho, \mathcal{C}, \kappa, \widehat{\kappa}, \alpha \rangle$ with resource and capacity constraints, if there exists a policy $\pi \in \Pi^{\text{HR}}$ that is a feasible solution for $M$, there exists a stationary deterministic policy $\pi^{\text{SD}} \in \Pi^{\text{SD}}$ that is also feasible, and the expected total reward of $\pi^{\text{SD}}$ is no less than that of $\pi$:*

$$\forall \; \pi \in \Pi^{\text{HR}}, \; \exists \; \pi^{\text{SD}} \in \Pi^{\text{SD}} : U_\gamma(\pi^{\text{SD}}, \alpha) \geq U_\gamma(\pi, \alpha)$$

**Proof:** See Appendix A.2. □

The result of Theorem 2 is not at all surprising: intuitively, stationary deterministic policies are optimal, because history dependence does not increase the utility of the policy, and using randomization can only increase resource costs. The latter is true because including an action in a policy incurs the same costs in terms of resources regardless of the probability of executing that action (or the expected number of times the action will be executed). This is true because we are dealing with non-consumable resources; this property does not hold for MDPs with consumable resources (as we discuss in more detail in Section 6).

We now show that uniformly optimal policies do not always exist for our constrained problem. This result is well known for another class of constrained MDPs, where constraints are imposed on the total *expected* costs that are proportional to the expected number of times the corresponding actions are executed (discussed earlier in Section 2). MDPs with such constraints arise, for example, when bounds are imposed on the expected usage of consumable resources, and as mentioned in Section 2, these problems can be solved using linear programming by augmenting the dual LP (6) with linear constraints on expected costs (8). Below, we establish the same result for problems with non-consumable resources and capacity constraints.

**Observation 1** *There do not always exist uniformly optimal solutions to (9). In other words, there exist two constrained MDPs that differ only in their initial conditions: $M = \langle \mathcal{S}, \mathcal{A}, p, r, \mathcal{O}, \rho, \mathcal{C}, \kappa, \widehat{\kappa}, \alpha \rangle$ and $M' = \langle \mathcal{S}, \mathcal{A}, p, r, \mathcal{O}, \rho, \mathcal{C}, \kappa, \widehat{\kappa}, \alpha' \rangle$, such that there is no policy that is optimal for both problems simultaneously, i.e., for any two policies $\pi$ and $\pi'$ that are optimal solutions to $M$ and $M'$, respectively, the following holds:*

$$U_\gamma(\pi, \alpha) > U_\gamma(\pi', \alpha), \quad U_\gamma(\pi, \alpha') < U_\gamma(\pi', \alpha') \quad (10)$$

We demonstrate this observation by example.

**Example 3** *Consider the resource-constrained problem as in Example 2. It is easy to see that if the initial conditions are $\alpha = [1, 0, 0]$ (the agent starts in state $s_1$ with certainty), the optimal policy for states $s_1$ and $s_2$ is the same as in Example 1 ($s_1 \to a_2$ and $s_2 \to a_3$), which, given the initial conditions, results in zero probability of reaching state $s_3$ (to which the noop $a_0$ is assigned). This policy requires the truck and the forklift. However, if the agent starts in state $s_3$ ($\alpha = [0, 0, 1]$), the optimal policy is to fix the truck (execute $a_4$ in $s_3$), and to resort to furniture delivery (do $a_1$ in $s_1$ and assign the noop $a_o$ to $s_2$, which is then never visited). This policy requires the mechanic and the truck. These two policies are uniquely optimal for the corresponding initial conditions, and are suboptimal for other initial conditions, which demonstrates that no uniformly optimal policy exists for this example.* $\square$

The intuition behind the fact that uniformly optimal policies do not, in general, exist for constrained MDPs is that since the resource information is not a part of the MDP state space, and there are constraints imposed on resource usage, the principle of Bellman optimality does not hold (optimal actions for different states cannot be chosen independently). Given a constrained MDP, it is possible to construct an equivalent unconstrained MDP with the standard properties of optimal solutions (by folding the resource information into the state space, and modeling resource constraints via the transition function), but the resulting state space will be exponential in the number of resources.

We now analyze the computational complexity of the optimization problem (9).

**Theorem 3** *The following decision problem is NP-complete. Given an instance of an MDP $\langle \mathcal{S}, \mathcal{A}, p, r, \mathcal{O}, \rho, \mathcal{C}, \kappa, \widehat{\kappa}, \alpha \rangle$ with resources and capacity constraints, and a rational number $Y$, does there exist a feasible policy $\pi$, whose expected total reward, given $\alpha$, is no less than $Y$?*

**Proof:** See Appendix A.3. □

Note that the above complexity result stems from the limited capacities of the agents and the fact that we define the resource requirements of a policy as the set of all resources needed to carry out all actions that have a nonzero probability of being executed. If, however, we defined constraints on the *expected* resource requirements, then actions with low probability of being executed would have lower resource requirements, optimal policies would be randomized, and the problem would be equivalent to a knapsack with continuously divisible items, which is solvable in polynomial time via the LP formulation of MDPs with linear constraints (6,8).

## 3.2 MILP Solution

Now that we have analyzed the properties of the optimization problem (9), we present a formulation of (9) as a mixed integer linear program (MILP). Given that we have established NP-completeness of (9) in the previous section, MILP (also NP-complete) is a reasonable formulation that allows us to reap the benefits of a vast selection of efficient algorithms and tools (see, for example, the text by Wolsey, 1998 and references therein).

In this section and in the rest of the paper we will assume that the resource requirements of actions are binary, i.e., $\rho(a, o) = \{0, 1\}$. We make this assumption to simplify the discussion, and it does not limit the generality of our results. We briefly describe the case of non-binary resource costs in Appendix B for completeness, but refer to the work by Dolgov (2006) for a more detailed discussion and examples.

Let us rewrite (9) in the occupation measure coordinates $x$ by adding the constraints from (9) to the standard LP in occupancy coordinates (6). Noticing that (for states with nonzero probability of being visited) $\pi(s, a)$ and $x(s, a)$ are either zero or nonzero simultaneously:

$$H\Big(\sum_s \pi(s, a)\Big) = H\Big(\sum_s x(s, a)\Big), \qquad \forall a \in \mathcal{A},$$

and that, when $\rho(a, o) = \{0, 1\}$, the total resource requirements of a policy can be simplified as follows:

$$\max_a \Big\{ \rho(a, o) H\Big(\sum_s x(s, a)\Big) \Big\} = H\Big(\sum_a \rho(a, o) \sum_s x(s, a)\Big), \qquad \forall o \in \mathcal{O}, \qquad (11)$$

we get the following program in $x$:

$$\max_x \sum_s \sum_a x(s, a) r(s, a)$$

subject to:

$$\sum_a x(\sigma, a) - \gamma \sum_s \sum_a x(s, a) p(\sigma | s, a) = \alpha(\sigma), \qquad \forall \sigma \in \mathcal{S};$$

$$\sum_o \kappa(o, c) H\Big(\sum_a \rho(a, o) \sum_s x(s, a)\Big) \leq \widehat{\kappa}(c), \qquad \forall c \in \mathcal{C}; \qquad (12)$$

$$x(s, a) \geq 0, \qquad \forall s \in \mathcal{S}, a \in \mathcal{A}.$$

The challenge in solving this mathematical program is that the constraints are nonlinear due to the Heaviside function $H$.

To linearize the Heaviside function, we augment the original optimization variables $x$ with a set of $|\mathcal{O}|$ binary variables $\delta(o) \in \{0, 1\}$, where $\delta(o) = H\left(\sum_a \rho(a, o) \sum_s x(s, a)\right)$. In other words, $\delta(o)$ is an indicator variable that shows whether the policy requires resource $o$. Using $\delta(o)$, we can rewrite the resource constraints in (12) as:

$$\sum_o \kappa(o, c)\delta(o) \leq \widehat{\kappa}(c), \qquad \forall c \in \mathcal{C}, \tag{13}$$

which are linear in $\delta$. We can then synchronize $\delta$ and $x$ via the following linear inequalities:

$$1/X \sum_a \rho(a, o) \sum_s x(s, a) \leq \delta(o), \qquad \forall o \in \mathcal{O}, \tag{14}$$

where $X \geq \max_o \sum_a \rho(a, o) \sum_s x(s, a)$ is the normalization constant, for which any upper bound on the argument of $H()$ can be used. The bound $X$ is guaranteed to exist for discounted problems. For example, we can use $X = (1 - \gamma)^{-1} \max_o \sum_a \rho(a, o)$, since $\sum_{s,a} x(s, a) = (1 - \gamma)^{-1}$ for any $x$ that is a valid occupation measure for an MDP with discount factor $\gamma$.[6]

Putting it all together, the problem (9) of finding optimal policies under resource constraints can be formulated as the following MILP:

$$\max_{x, \delta} \sum_s \sum_a x(s, a)r(s, a)$$

subject to:

$$\sum_a x(\sigma, a) - \gamma \sum_s \sum_a x(s, a)p(\sigma|s, a) = \alpha(\sigma), \qquad \forall \sigma \in \mathcal{S};$$

$$\sum_o \kappa(o, c)\delta(o) \leq \widehat{\kappa}(c), \qquad \forall c \in \mathcal{C}; \tag{15}$$

$$1/X \sum_a \rho(a, o) \sum_s x(s, a) \leq \delta(o), \qquad \forall o \in \mathcal{O};$$

$$x(s, a) \geq 0, \qquad \forall s \in \mathcal{S}, a \in \mathcal{A};$$

$$\delta(o) \in \{0, 1\}, \qquad \forall o \in \mathcal{O}.$$

We illustrate the MILP construction with an example.

**Example 4** *Let us formulate the MILP for the constrained problem from Example 3. Recall that in that problem there are three resources $\mathcal{O} = \{o_t, o_f, o_m\}$ (truck, forklift, and mechanic), one capacity type $\mathcal{C} = \{c_1\}$ (money), and actions have the following resource requirements (again, listing only the nonzero ones):*

$$\rho(a_1, o_t) = 1, \ \rho(a_2, o_t) = 1, \ \rho(a_2, o_f) = 1, \ \rho(a_3, o_t) = 1, \ \rho(a_4, o_t) = 1, \ \rho(a_4, o_m) = 1$$

---

6. Instead of using a single $X$ for all resources, a different $X(o) \geq \sum_a \rho(a, o) \sum_s x(s, a)$ can be used for every resource, leading to more uniform normalization and potentially better numerical stability of the MILP solver.

*The resources have the following capacity costs:*

$$\kappa(o_t, c_1) = 2, \quad \kappa(o_f, c_1) = 3, \quad \kappa(o_m, c_1) = 4,$$

*and the agent has a limited budget, i.e., a capacity bound, $\hat{\kappa}(c_1) = 8$.*

*To compute the optimal policy for an arbitrary $\alpha$, we can formulate the problem as an MILP as described above. Using binary variables $\delta(o) = \{\delta(o_t), \delta(o_f), \delta(o_m)\}$, we can express the constraint on capacity cost as the following inequality:*

$$2\delta(o_t) + 3\delta(o_f) + 4\delta(o_m) \leq 8,$$

*For the constraints that synchronize the occupation measure $x$ and the binary indicators $\delta(o)$, we can set $X = (1-\gamma)^{-1} \max_o \sum_a \rho(a, o) = 4(1-\gamma)^{-1}$. Combining this with other constraints from (15), we get an MILP with 12 continuous and 4 binary variables, and $|\mathcal{S}| + |\mathcal{C}| + |\mathcal{O}| = 3 + 3 + 1 = 7$ constraints (not counting the last two sets of range constraints).* $\square$

As mentioned earlier, even though solving such programs is, in general, an NP-complete task, there is a wide variety of very efficient algorithms and tools for doing so. Therefore, one of the benefits of formulating the optimization problem (9) as an MILP is that it allows us to make use of the highly efficient existing tools.

## 4. Multiagent Resource Allocation

We now consider the multiagent problem of resource allocation between several agents, where the resource preferences of the agents are defined by the constrained MDP model described in the previous section. We reiterate our main assumptions about the problem:

1. **Weak coupling.** We assume that agents are weakly-coupled (Meuleau et al., 1998), i.e., they only interact through the shared resources, and once the resources are allocated, the agents' transitions and rewards are independent. This assumption is critical to our results.[7]

2. **One-shot resource allocation.** The resources are distributed once before the agents start executing their MDPs. There is no reallocation of resources during the MDP phase. This assumption is critical to our results; allowing reallocation of resources would violate the weak-coupling assumption.

3. **Initial central control over the resources.** We assume that at the beginning of the resource-allocation phase, the resources are controlled by a single authority. This is the standard sell-auction setting. For problems where the resources are distributed among the agents to begin with, we face the problem of designing a computationally-efficient *combinatorial exchange* (Parkes, Kalagnanam, & Eso, 2001), which is a more complicated problem that is outside the scope of this work. However, many of the ideas presented in this paper could potentially be applicable to that domain as well.

---

7. If agents are cooperative, the assumption about weak coupling can be relaxed (at the expense of an increase in complexity), and the same MILP-based algorithm for simultaneously performing policy optimization and resource allocation can be applied if we consider the joint state spaces of the interacting agents. For self-interested agents, a violation of the weakly-coupled assumption would mean that the agents would be playing a Markov game (Shapley, 1953) once the resources are allocated, which would significantly complicate the strategic analysis of the agents' bidding strategies during the initial resource allocation.

4. **Binary resource costs.** As before, we assume that agents' resource costs are binary. This assumption is not limiting. The case of non-binary resources is discussed in Appendix B.

Formally, the input to the resource-allocation problem consists of the following:

- $\mathcal{M}$ is the set of agents; we will use $m \in \mathcal{M}$ to refer to an agent.

- $\{\langle \mathcal{S}, \mathcal{A}, p^m, r^m, \alpha^m, \rho^m, \widehat{\kappa}^m \rangle\}$ is the collection of weakly-coupled single-agent MDPs, as defined in the single-agent model in Section 3. For simplicity, but without loss of generality, we assume that all agents have the same state and action spaces $\mathcal{S}$ and $\mathcal{A}$, but each has its own transition and reward functions $p^m$ and $r^m$, initial conditions $\alpha^m$, as well as its own resource requirements $\rho^m : \mathcal{A} \times \mathcal{O} \mapsto \{0, 1\}$ and capacity bounds $\widehat{\kappa}^m : \mathcal{C} \mapsto \mathbb{R}$. We also assume that all agents have the same discount factor $\gamma$, but this assumption can be trivially relaxed.

- $\mathcal{O}$ and $\mathcal{C}$ are the sets of resources and capacities, defined exactly as in the single-agent model in Section 3.

- $\kappa : \mathcal{O} \times \mathcal{C} \mapsto \mathbb{R}$ specifies the capacity costs of the resources, defined exactly as in the single-agent model in Section 3.

- $\widehat{\rho} : \mathcal{O} \mapsto \mathbb{R}$ specifies the upper bound on the amounts of the shared resources (this defines the additional bound for the multiagent problem).

Given the above, our goal is to design a mechanism for allocating the resources to the agents in an economically efficient way, i.e., in a way that maximizes the social welfare of the agents (one of the most often-used criteria in mechanism design). We would also like the mechanism to be efficient from the computational standpoint.

**Example 5** *Suppose that there are two delivery agents. The MDP and capacity constraints of the first agent are exactly as they were defined previously in Examples 1 and 2. The MDP of the second agent is almost the same as that of the first agent, with the only difference that it gets a slightly higher reward for delivering appliances: $r^2(s_1, a_2) = 12$ (whereas $r^1(s_1, a_2) = 10$ for the first agent). Suppose there are two trucks, one forklift, and one mechanic that are shared by the two agents. These bounds are specified as follows:*

$$\widehat{\rho}(o_t) = 2, \quad \widehat{\rho}(o_f) = 1, \quad \widehat{\rho}(o_m) = 1.$$

*Then the problem is to decide which agent should get the forklift, and which should get the mechanic (trucks are plentiful in this example).* □

## 4.1 Combinatorial Auctions

As previously mentioned, the problem of finding an optimal resource allocation among self-interested agents that have complex valuations over combinations of resources arises in many different domains (e.g., Ferguson, Nikolaou, Sairamesh, & Yemini, 1996; Wellman et al., 2001) and is often called a *combinatorial allocation problem*. A natural and widely used mechanism for solving such problems is a *combinatorial auction* (CA) (e.g., de Vries & Vohra, 2003). In a CA, each agent submits a set of bids for resource bundles to the auctioneer, who then decides what resources each agent will get and at what price.

Consider the problem of allocating among a set of agents $\mathcal{M}$ a set of indivisible resources $\mathcal{O}$, where the total quantity of resource $o \in \mathcal{O}$ is bounded by $\hat{\rho}(o)$. Our earlier simplifying assumption that actions' resource requirements are binary implies that agents will only be interested in bundles that contain no more than one unit of a particular resource.

In a combinatorial auction, each agent $m \in \mathcal{M}$ submits a bid $b_w^m$ (specifying how much the agent is willing to pay) for every bundle $w \in \mathcal{W}^m$ that has some value $u_w^m > 0$. In some cases, it is possible to express such bids without enumerating all bundles (for example, using an XOR bidding language (Sandholm, 1999) where it is necessary to only consider bundles with strictly positive value, such that no subset of a bundle has the same value). Such techniques often reduce the complexity of the resource-allocation problem, but do not, in general, avoid the exponential blow up in the number of bids. Therefore, below we describe the simplest combinatorial auction with flat bids, but it should be noted that many concise bidding languages exist that in special cases can reduce the number of explicit bids.

After collecting all the bids, the auctioneer solves the winner-determination problem (WDP), a solution to which prescribes how the resources should be distributed among the agents and at what prices. If agent $m$ wins bundle $w$ at price $q_w^m$, its utility is $u_w^m - q_w^m$ (we are assuming risk-neutral agents with quasi-linear utility functions). Thus, the optimal bidding strategy of an agent depends on how the auctioneer allocates the resources and sets prices.

Vickrey-Clarke-Groves (VCG) mechanisms (Vickrey, 1961; Clarke, 1971; Groves, 1973) are a widely used family of mechanisms that have certain very attractive properties (discussed in more detail below). An instantiation of a VCG mechanism in the context of combinatorial auctions is the Generalized Vickrey Auction (GVA) (MacKie-Mason & Varian, 1994), which allocates resources and sets prices as follows. Given the bids $b_w^m$ of all agents, the auctioneer chooses an allocation that maximizes the sum of agents' bids. This problem is NP-complete (Rothkopf et al., 1998) and can be expressed as the following integer program, where the optimization variables $z_w^m = \{0, 1\}$ are indicator variables that show whether bundle $w$ is assigned to agent $m$, and $n_{wo} = \{0, 1\}$ specifies whether bundle $w$ contains $o$:[8]

---

8. There are other related algorithms for solving the WDP (e.g., Sandholm, 2002), but we will use the integer program (16) as a representative formulation for the class of algorithms that perform a search in the space of binary decisions on resource bundles.

$$\max_z \sum_{m \in \mathcal{M}} \sum_{w \in \mathcal{W}^m} z_w^m b_w^m$$

subject to:

$$\sum_{w \in \mathcal{W}^m} z_w^m \leq 1, \qquad \forall m \in \mathcal{M};$$

$$\sum_{m \in \mathcal{M}} \sum_{w \in \mathcal{W}^m} z_w^m n_{wo} \leq \widehat{\rho}(o), \qquad \forall o \in \mathcal{O}.$$

(16)

The first constraint in (16) says that no agent can receive more than one bundle, while the second constraint ensures that the total amount of resource $o$ assigned to the agents does not exceed the total amount available. Notice that MILP (16) performs the summation over exponentially large sets of bundles $w \in \mathcal{W}^m$. As outlined above, in an auction with XOR bidding, these sets would typically be smaller, but, in general, still exponentially large.

A GVA assigns resources according to the optimal solution $\widetilde{z}$ to (16) and sets the payment for agent $m$ to:

$$q_w^m = V_{-m}^* - \sum_{m' \neq m} \widetilde{z}_w^{m'} b_w^{m'},$$

(17)

where $V_{-m}^*$ is the value of (16) if $m$ were to not participate in the auction (the optimal value if $m$ does not submit any bids), and the second term is the sum of other agents' bids in the solution $\widetilde{z}$ to the WDP with $m$ participating.

A GVA has a number of nice properties. It is *strategy-proof*, meaning that the dominant strategy of every agent is to bid its true value for every bundle: $b_w^m = u_w^m$. The auction is *economically efficient*, meaning that it allocates the resources to maximize the social welfare of the agents (because, when agents bid their true values, the objective function of (16) becomes the social welfare). Finally, a GVA satisfies the *participation constraint*, meaning that no agent decreases its utility by participating in the auction.

A straightforward way to implement a GVA for our MDP-based problem is the following. Let each agent $m \in \mathcal{M}$ enumerate all resource bundles $\mathcal{W}^m$ that satisfy its local capacity constraints defined by $\widehat{\kappa}^m(c)$ (this is sufficient because our MDP model implies free disposal of resources for the agents, and we make the same assumption about the auctioneer). For each bundle $w \in \mathcal{W}^m$, agent $m$ would determine the feasible action set $\mathcal{A}(w)$ and formulate an MDP $\Lambda^m(w) = \langle \mathcal{S}, \mathcal{A}(w), p^m(w), r^m(w), \alpha^m \rangle$, where $p^m(w)$ and $r^m(w)$ are the transition and reward functions defined on the pruned action space $\mathcal{A}(w)$. Every agent would then solve each $\Lambda^m(w)$ corresponding to a feasible bundle to find the optimal policy $\widetilde{\pi}^m(w)$, whose expected discounted reward would define the value of bundle $w$: $u_w^m = U_\gamma^m(\widetilde{\pi}^m(w), \alpha^m)$.

This mechanism suffers from two major complexity problems. First, the agents have to enumerate an exponential number of resource bundles and compute the value of each by solving the corresponding (possibly large) MDP. Second, the auctioneer has to solve an NP-complete winner-determination problem on exponentially large input. The following sections are devoted to tackling these complexity problems.

**Example 6** *Consider the two-agent problem described in Example 5, where two trucks, one forklift, and the services of one mechanic are being auctioned off. Using the straightforward version of the combinatorial auction outlined above, each agent would have to consider* $2^{|\mathcal{O}|} = 2^3 = 8$ *possible resource bundles (since resource requirements of both agents are binary, neither agent is going to bid on a bundle that contains two trucks). For every resource bundle, each agent will have to formulate and solve the corresponding MDP to compute the utility of the bundle.*

*For example, if we assume that both agents start in state $s_1$ (different initial conditions would result in different expected rewards, and thus different utility functions), the value of the null resource bundle to both agents would be 0 (since the only action they would be able to execute is the noop $a_0$). On the other hand, the value of bundle $[o_t, o_f, o_m] = [1, 1, 1]$ that contains all the resources would be 95.3 to the first agent and 112.4 to the second one. The value of bundle $[1, 1, 0]$ to each agent would be the same as the value of $[1, 1, 1]$ (since their optimal policies for the initial conditions that put them in $s_1$ do not require the mechanic).*

*Once the agents submit their bids to the auctioneer, it will have to solve the WDP via the integer program (16) with $|\mathcal{M}|2^{|\mathcal{O}|} = 2(2)^3 = 16$ binary variables. Given the above, the optimal way to allocate the resources would be to assign a truck to each of the agents, the forklift to the second agent, and the mechanic to either (or neither) of the two. Thus, the agents would receive bundles $[1, 0, 0]$ and $[1, 1, 0]$, respectively, resulting in social welfare of $50 + 112.4 = 162.4$. However, if at least one of the agents had a non-zero probability of starting in state $s_3$, the value of the resource bundles involving the mechanic would change drastically, as would the optimal resource allocation and its social value.* □

## 4.2 Avoiding Bundle Enumeration

To avoid enumerating all resource bundles that have non-zero value to an agent, two things are required: i) the mechanism has to support a concise bidding language that allows the agent to express its preferences to the auctioneer in a compact manner, and ii) the agents have to be able to find a good representation of their preferences in that language. A simple way to achieve both in our model is to create an auction where the agents submit the specifications of their resource-parameterized MDPs to the auctioneer as bids: the language is compact and, given our assumption that each agent can formulate its planning problem as an MDP, this does not require additional computation for the agents. However, this only changes the communication protocol between the agents and the auctioneer, similarly to other concise bidding languages (Sandholm, 1999; Nisan, 2000; Boutilier & Hoos, 2001; Boutilier, 2002). As such, it simply moves the burden of solving the valuation problem from the agents to the auctioneer, which by itself does not lead to any gains in computational efficiency. Such a mechanism also has implications on information privacy issues, because the agents have to reveal their local MDPs to the auctioneer (which they might not want to do). Nevertheless, we can build on this idea to increase the efficiency of solving both the valuation and winner-determination problems and do so while keeping most of the agents' MDP information private. We address ways of maintaining information privacy in the next section, and for the moment focus on improving the computational complexity of the agents' valuation and the auctioneer's winner-determination problems.

The question we pose in this section is as follows. Given that the bid of each agent consists of its MDP, its resource information and its capacity bounds $\langle \mathcal{S}, \mathcal{A}, p^m, r^m, \alpha^m, \rho^m, \widehat{\kappa}^m \rangle$, can the auctioneer formulate and solve the winner-determination problem more efficiently than by simply enumerating each agent's resource bundles and solving the standard integer program (16) with an exponential number of binary variables?

Therefore, the goal of the auctioneer is to find a joint policy (a collection of single-agent policies under our weak-coupling assumption) that maximizes the sum of the expected total discounted rewards for all agents, under the conditions that: i) no agent $m$ is assigned a set of resources that violates its capacity bound $\widehat{\kappa}^m$ (i.e., no agent is assigned more resources than it can carry), and ii) the total amounts of resources assigned to all agents do not exceed the global resource bounds $\widehat{\rho}(o)$ (i.e., we cannot allocate to the agents more resources than are available). This problem can be expressed as the following mathematical program:

$$\max_{\pi} \sum_m U^m_\gamma(\pi^m, \alpha^m)$$

subject to:

$$\sum_o \kappa(o, c) H\big(\rho^m(a, o) \sum_s \pi^m(s, a)\big) \leq \widehat{\kappa}^m(c), \qquad \forall c \in \mathcal{C}, m \in \mathcal{M};$$

$$\sum_m \rho^m(a, o) H\big(\sum_s \pi^m(s, a)\big) \leq \widehat{\rho}(o), \qquad \forall o \in \mathcal{O}. \tag{18}$$

Obviously, the decision version of this problem is NP-complete, as it subsumes the single-agent MDP with capacity constraints, NP-completeness of which was shown in Section 3.1. Moreover, the problem remains NP-complete even in the absence of single-agent capacity constraints. Indeed, the global constraint on the amounts of the shared resources is sufficient to make the problem NP-complete, which can be shown with a straightforward reduction from KNAPSACK, similar to the one used in the single-agent case in Section 3.1.

We can linearize (18) similarly to the single-agent problem from Section 3.2, yielding the following MILP, which is simply a multiagent version of (15) (recall the assumption of this section that resource requirements are binary):

$$\max_{x, \delta} \sum_m \sum_s \sum_a x^m(s, a) r^m(s, a)$$

subject to:

$$\sum_a x^m(\sigma, a) - \gamma \sum_s \sum_a x^m(s, a) p^m(\sigma|s, a) = \alpha^m(\sigma), \qquad \forall \sigma \in \mathcal{S}, m \in \mathcal{M};$$

$$\sum_o \kappa(o, c) \delta^m(o) \leq \widehat{\kappa}^m(c), \qquad \forall c \in \mathcal{C}, m \in \mathcal{M};$$

$$\sum_m \delta^m(o) \leq \widehat{\rho}(o), \qquad \forall o \in \mathcal{O}; \tag{19}$$

$$1/X \sum_a \rho^m(a, o) \sum_s x^m(s, a) \leq \delta^m(o), \qquad \forall o \in \mathcal{O}, m \in \mathcal{M};$$

$$x^m(s, a) \geq 0, \qquad \forall s \in \mathcal{S}, a \in \mathcal{A}, m \in \mathcal{M};$$

$$\delta^m(o) \in \{0, 1\}, \qquad \forall o \in \mathcal{O}, m \in \mathcal{M},$$

where $X \geq \max_{o,m} \sum_a \rho(a, o) \sum_s x^m(s, a)$ is an upper bound on the argument of $H(\cdot)$, used for normalization. As in the single-agent case, this bound is guaranteed to exist for discounted MDPs and is easy to obtain.

The MILP (19) allows the auctioneer to solve the WDP without having to enumerate the possible resource bundles. As compared to the standard WDP formulation (16), which has on the order of $|\mathcal{M}|2^{|\mathcal{O}|}$ binary variables, (19) has only $|\mathcal{M}||\mathcal{O}|$ binary variables. This exponential reduction is attained by exploiting the knowledge of the agents' MDP-based valuations and simultaneously solving the policy-optimization and resource-allocation problems. Given that the worst-case solution time for MILPs is exponential in the number of integer variables, this reduction has a significant impact on the worst-case performance of the mechanism. The average-case running time is also reduced drastically, as demonstrated by our experiments, presented in Section 5.

**Example 7** *If we apply the mechanism discussed above to our running example as an alternative to the straightforward combinatorial auction presented in Example 6, the winner-determination MILP (19) will look as follows. It will have $|\mathcal{M}||\mathcal{S}||\mathcal{A}| = (2)(3)(5) = 30$ continuous occupation-measure variables $x^m$, and $|\mathcal{M}||\mathcal{O}| = (2)(3) = 6$ binary variables $\delta^m(o)$. It will have $|\mathcal{M}||\mathcal{S}| = (2)(3) = 6$ conservation-of-flow constraints that involve continuous variables only, as well as $|\mathcal{M}||\mathcal{C}| + |\mathcal{O}| + |\mathcal{M}||\mathcal{O}| = (2)(1) + 3 + (2)(3) = 9$ constraints that involve binary variables.*

*The capacity constraints for the agents will be exactly as in the single-agent case described in Example 4, and the global resource constraints will be:*

$$\delta^1(o_t) + \delta^2(o_t) \le 2, \qquad \delta^1(o_f) + \delta^2(o_f) \le 1, \qquad \delta^1(o_m) + \delta^2(o_m) \le 1.$$

*Notice that in this example there is one binary decision variable per resource per agent (yielding 6 such variables for this simple problem). This is exponentially fewer than the number of binary variables in the straightforward CA formulation of Example 6, which requires one binary variable per resource bundle per agent (yielding 16 such variables for this problem). Given that MILPs are NP-complete in the number of integer variables, this reduction from 16 to 6 variables is noticeable even in a small problem like this one and can lead to drastic speedup for larger domains.* □

The mechanism described above is an instantiation of the GVA, so by the well-known properties of VCG mechanisms, this auction is strategy-proof (the agents have no incentive to lie to the auctioneer about their MDPs), it attains the socially optimal resource allocation, and no agent decreases its utility by participating in the auction.

To sum up the results of this section: by having the agents submit their MDP information to the auctioneer instead of their valuations over resource bundles, we have essentially removed all computational burden from the agents and at the same time significantly simplified the auctioneer's winner-determination problem (the number of integer variables in the WDP is reduced exponentially).

### 4.3 Distributing the Winner-Determination Problem

Unlike the straightforward combinatorial auction implementation discussed earlier in Section 4.1, where the agents shared some computational burden with the auctioneer, in the mechanism from Section 4.2, the agents submit their information to the auctioneer and then just idle while waiting for a solution. This suggests further potential improvements in computational efficiency. Indeed, given the complexity of MILPs, it would be beneficial to exploit the computational power of the agents to offload some of the computation from the auctioneer back to the agents (we assume that agents have no cost for "helping out" and would prefer for the outcome to be computed faster).[9] Thus, we would like to distribute the computation of the winner-determination problem (19), the common objective in *distributed algorithmic mechanism design* (Feigenbaum & Shenker, 2002; Parkes & Shneidman, 2004).[10]

For concreteness, we will base the algorithm of this section on the branch and bound method for solving MILPs (Wolsey, 1998), but exactly the same techniques will also work for other MILP algorithms (e.g., cutting planes) that perform a search in the space of *LP relaxations* of the MILP. In branch and bound for MILPs with binary variables, LP relaxations are created by choosing a binary variable and setting it to either 0 or 1, and relaxing the integrality constraints of other binary variables. If a solution to an LP relaxation happens to be integer-valued, it provides a lower bound on the value of the global solution. A non-integer solution provides an upper bound for the current subproblem, which (combined with other lower bounds) is used to prune the search space.

Thus, a simple way for the auctioneer to distribute the branch and bound algorithm is to simply farm out LP relaxations to other agents and ask them to solve the LPs. However, it is easy to see that this mechanism is not strategy-proof. Indeed, an agent that is tasked with performing some computation for determining the optimal resource allocation or the associated payments could benefit from lying about the outcome of its computation to the auctioneer. This is a common phenomenon in distributed mechanism implementations: whenever some WDP calculations are offloaded to an agent participating in the auction, the agent might be able to benefit from sabotaging the computation. There are several methods to ensuring the strategy-proofness of a distributed implementation. The approach best suited for our problem is based on the idea of *redundant computation* (Parkes & Shneidman, 2004),[11] where multiple agents are asked to do the same task and any disagreement is carefully punished to discourage lying. In the rest of this section, we demonstrate that this is very easy to implement in our case.

The basic idea is very simple: let the auctioneer distribute LP relaxations to the agents, but check their solutions and re-solve the problems if the agents return incorrect solutions (this would make truthful computation a weakly-dominant strategy for the agents, and a nonzero punishment can be used to achieve strong dominance). This strategy of the auctioneer removes the incentive for the agents to lie and yields exactly the same solution as the centralized algorithm. However, in order for this to be beneficial, the complexity of checking a solution must be significantly lower than the complexity of solving the problem. Fortunately, this is true for LPs.

Suppose the auctioneer has to solve the following LP, which can be written in two equivalent ways (let us refer to the one on the left as the primal, and to the one on the right

---

9. As observed by Parkes and Shneidman (2004), this assumption is a bit controversial, since a desire for efficient computation implies nonzero cost for computation, while the agents' cost for "helping out" is not modeled. It is, nonetheless, a common assumption in distributed mechanism implementations.

10. We describe one simple way of distributing the mechanism, others are also possible.

11. Redundant computation is discussed by Parkes and Shneidman (2004) in the context of *ex post* Nash equilibria, whereas we are interested in dominant strategies, but the high-level idea is very similar.

as the dual):

$$\min \boldsymbol{\alpha}^T \mathbf{v} \qquad\qquad \max \mathbf{r}^T \mathbf{x}$$
$$\text{subject to:} \qquad\qquad \text{subject to:} \qquad\qquad (20)$$
$$A^T \mathbf{v} \geq \mathbf{r}. \qquad\qquad A\mathbf{x} = \boldsymbol{\alpha}; \quad \mathbf{x} \geq 0.$$

By the strong duality property, if the primal LP has a solution $\mathbf{v}^*$, then the dual also has a solution $\mathbf{x}^*$, and $\boldsymbol{\alpha}^T \mathbf{v}^* = \mathbf{r}^T \mathbf{x}^*$. Furthermore, given a solution to the primal LP, it is easy to compute a solution to the dual: by complementary slackness, $\mathbf{v}^{*T} = \mathbf{r}^T B^{-1}$ and $\mathbf{x}^* = B^{-1}\boldsymbol{\alpha}$, where $B$ is a square invertible matrix composed of columns of $A$ that correspond to basic variables of the solution.

These well-known properties can be used by the auctioneer to quickly check optimality of solutions returned by the agents. Suppose that an agent returns $\mathbf{v}$ as a solution to the primal LP. The auctioneer can calculate the dual solution $\mathbf{v}^T = \mathbf{r}^T B^{-1}$ and check whether $\mathbf{r}^T \mathbf{x} = \boldsymbol{\alpha}^T \mathbf{v}$. Thus, the most expensive operation that the auctioneer has to perform is the inversion of $B$, which can be done in sub-cubic time. As a matter of fact, from the implementation perspective, it would be more efficient to ask the agents to return both the primal and the dual solutions, since many popular algorithms compute both in the process of solving LPs.

Thus, we have provided a simple method that allows us to effectively distribute the winner-determination problem, while maintaining strategy-proofness of the mechanism and with a negligible computation overhead for the auctioneer.

### 4.4 Preserving Information Privacy

The mechanism that we have discussed so far has the drawback that it requires agents to reveal complete information about their MDPs to the auctioneer. The problem is also exacerbated in the distributed WDP algorithm from the previous section, since not only does each agent reveal its MDP information to the auctioneer, but that information is then also spread to other agents via the LP relaxations of the global MILP. We now show how to alleviate this problem.

Let us note that, in saying that agents prefer not to reveal their local information, we are implicitly assuming that there is an external factor that affects agents' utilities that is not captured in the agents' MDPs. A sensible way to measure the value of information is by how it changes one's decision-making process and its outcomes. Since this effect is not part of our model (in fact, it contradicts our weak-coupling assumption), we cannot in a domain-independent manner define what constitutes "useful" information, and how bad it is for an agent to reveal too much about its MDP. Modeling such effects and carefully analyzing them is an interesting research task, but it is outside the scope of this paper. Thus, for the purposes of this section, we will be content with a mechanism that hides enough information to make it impossible for the auctioneer or an agent to uniquely determine the transition or reward function of any other agent (in fact, information revealed to any agent will map to infinitely many MDPs of other agents).[12] Many such transformations are possible; we present just one to illustrate the concept.

---

12. A more stringent condition would require that agents' preferences over resource bundles are not revealed (Parkes, 2001), but we set a lower bar here.

The main idea of our approach is to modify the previous mechanism so that the agents submit their private information to the auctioneer in an "encrypted" form that allows the auctioneer to solve the winner-determination problem, but does not allow it to infer the agents' original MDPs.

First, note that, instead of passing an MDP to the auctioneer, each agent can submit an equivalent LP (6). So, the question becomes: can the agent transform its LP in such a way that the auctioneer will be able to solve it, but will not be able to infer the transition and reward functions of the originating MDP? In other words, the problem reduces to the following. Given an LP $\mathcal{L}_1$ (created from an MDP $\Lambda = \langle \mathcal{S}, \mathcal{A}, p, r, \alpha \rangle$ via (6)), we need to find a transformation $\mathcal{L}_1 \to \mathcal{L}_2$ such that a solution to the transformed LP $\mathcal{L}_2$ will uniquely map to a solution to the original LP $\mathcal{L}_1$, but $\mathcal{L}_2$ will not reveal the transition or the reward functions of the original MDP ($p$ or $r$). We show that a simple change of variables suffices.

Suppose agent $m_1$ has an MDP-originated LP and is going to ask agent $m_2$ to solve it. In order to maintain the linearity of the problem (to keep it simple for $m_2$ to solve), $m_1$ should limit itself to linear transformations. Consider a linear, invertible transformation of the primal coordinates $\mathbf{u} = F\mathbf{v}$, and a linear, invertible transformation of the dual coordinates $\mathbf{y} = D\mathbf{x}$. Then, the LP from (20) will be transformed (by applying $F$, switching to the dual, and then applying $D$) to an equivalent LP in the new coordinates $\mathbf{y}$:

$$
\begin{aligned}
&\max \mathbf{r}^T D^{-1} \mathbf{y} \\
&\text{subject to:} \\
&(F^{-1})^T A D^{-1} \mathbf{y} = (F^{-1})^T \boldsymbol{\alpha}; \\
&D^{-1} \mathbf{y} \ge 0.
\end{aligned}
\tag{21}
$$

The value of the optimal solution to (21) will be the same as the value of the optimal solution to (20), and given an optimal solution $\mathbf{y}^*$ to (21), it is easy to compute the solution to the original: $\mathbf{x}^* = D^{-1}\mathbf{y}^*$. Indeed, from the perspective of the dual, the primal transformation $F$ is equivalent to a linear transformation of the dual equality constraints $A\mathbf{x} = \alpha$, which (given that $F$ is non-singular) has no effect on the solution or the objective function. Furthermore, the dual transformation $D$ is equivalent to a change of variables that modifies the solution but not the value of the objective function.

However, a problem with the above transformations is that it gives away $D^{-1}$. Indeed, agent $m_2$ will be able to simply read (up to a set of multiplicative constants) the transformation off the constraints $D^{-1}\mathbf{y} \ge 0$. Therefore, only diagonal matrices with positive coefficients (which are equivalent to stretching the coordinate system) are not trivially deduced by $m_2$, since they also map to $\mathbf{y} \ge 0$. Choosing a negative multiplier for some $x_i$ (inverting the axis) is pointless, because that flips the non-negativity constraints to $y_i \le 0$, immediately revealing the sign to $m_2$.

Let us demonstrate that, given any MDP $\Lambda$ and the corresponding LP $\mathcal{L}_1$, we can choose $D$ and $F$ such that it will be impossible for $m_2$ to determine the coefficients of $\mathcal{L}_1$ (or equivalently the original transition and reward functions $p$ and $r$). When agent $m_2$ receives $\mathcal{L}_2$ (as in (21)), all it knows is that $\mathcal{L}_2$ was created from an MDP, so the columns of the constraint matrix of the original LP $\mathcal{L}_1$ must sum to a constant:

$$
\sum_j A_{ji} = 1 - \gamma \sum_\sigma p(\sigma | s, a) = 1 - \gamma.
\tag{22}
$$

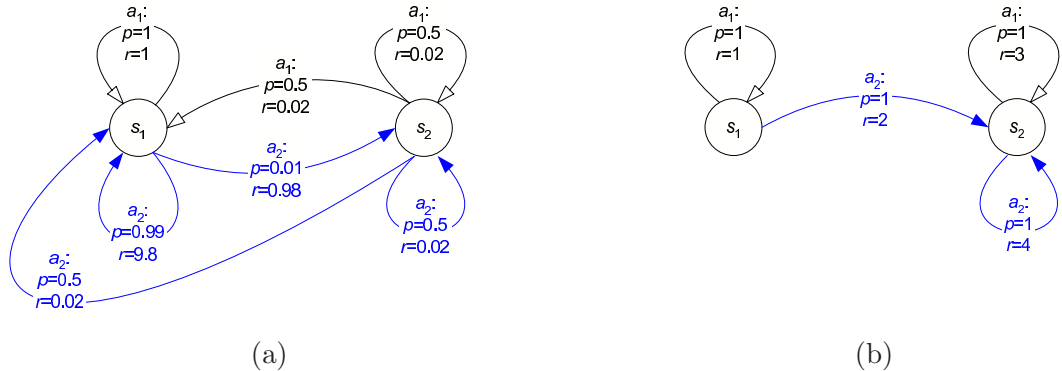

(a)                                                        (b)

Figure 3: Preserving privacy example. Two different MDPs that can lead to the same LP constraint matrix.

This gives $m_2$ a system of $|\mathcal{S}|$ nonlinear equations for the diagonal $D$ and arbitrary $F$, which have a total of $|\mathcal{S}||\mathcal{A}| + |\mathcal{S}|^2$ free parameters. For everything but the most degenerate cases (which can be easily handled by an appropriate choice of $D$ and $F$), these equations are hugely under-constrained and will have infinitely many solutions. As a matter of fact, by sacrificing $|\mathcal{S}|$ of the free parameters, $m_1$ can choose $D$ and $F$ in such a way that the columns of constraints in $\mathcal{L}_2$ will also sum to a constant $\gamma' \in (0, 1)$, which would have the effect of transforming $\mathcal{L}_1$ to an $\mathcal{L}_2$ that corresponds to another valid MDP $\Lambda_2$. Therefore, given an $\mathcal{L}_2$, there are infinitely many original MDPs $\Lambda$ and transformations $D$ and $F$ that map to the same LP $\mathcal{L}_2$.

We also have to consider the connection of resource and capacity costs to agents' occupation measures in the global WDP (19). There are two things that the auctioneer has to be able to do: i) determine the value of each agent's policy (to be able to maximize the social welfare), and ii) determine the resource requirements of the policies (to check the resource constraints). So, the question is, how does our transformation affect these? As noted earlier, the transformation does not change the objective function, so the first requirement holds. On the other hand, $D$ does change the occupation measure $x^m(s, a)$ by arbitrary multipliers. However, a multiplicative factor of $x^m(s, a)$ has no effect on the usage of non-consumable resources, as it only matters whether the corresponding $x^m(s, a)$ is zero or not (step function $H$ nullifies the scaling effect). Thus, the second condition also holds.

**Example 8** *Consider the two-state MDP depicted in Figure 3a that represents the decision-making problem of a sales company, with the two states corresponding to possible market conditions, and the two actions — to two possible sales strategies. Market conditions in state $s_1$ are much more favorable than in state $s_2$ (the rewards for both actions are higher). The transitions between the two states correspond to probabilities of market conditions changing and the rewards reflect the expected profitability in these two states. Obtaining such numbers in a realistic scenario would require performing costly and time-consuming research, and the company might not want to make this information public.*

*Therefore, if the company were to participate in the resource-allocation mechanism described above, it would want to encrypt its MDP before submitting it to the auctioneer.*

The MDP has the following reward function

$$r = (1, \ 19.622, \ 0.063, \ 0.084)^T, \tag{23}$$

and the following transition function:

$$p(a_1) = \begin{pmatrix} 1 & 0 \\ 0.5 & 0.5 \end{pmatrix}, \quad p(a_2) = \begin{pmatrix} 0.986 & 0.014 \\ 0.5 & 0.5 \end{pmatrix}. \tag{24}$$

Using $\gamma = 0.8$, this corresponds to the following conservation of flow constraint matrix:

$$A = \begin{pmatrix} 0.2 & 0.212 & -0.4 & -0.4 \\ 0 & -0.012 & 0.6 & 0.6 \end{pmatrix}. \tag{25}$$

Before submitting its LP to the auctioneer, the agent applies the following transformations:

$$D = diag(1, \ 0.102, \ 47.619, \ 47.619), \quad F = \begin{pmatrix} 2 & 0 \\ -0.084 & 0.126 \end{pmatrix}, \tag{26}$$

yielding the following new constraint matrix:

$$A' = (F^{-1})^T A D^{-1} = \begin{pmatrix} 0.1 & 1 & 0 & 0 \\ 0 & -0.9 & 0.1 & 0.1 \end{pmatrix}. \tag{27}$$

However, the above constraint matrix $A'$ corresponds to a non-transformed conservation of flow constraint for a different MDP (shown in Figure 3b) with $\gamma = 0.9$, the following reward function:

$$r = (1, \ 2, \ 3, \ 4)^T, \tag{28}$$

and the following transition function:

$$p(a_1) = \begin{pmatrix} 1 & 0 \\ 0 & 1 \end{pmatrix}, \quad p(a_2) = \begin{pmatrix} 0 & 1 \\ 0 & 1 \end{pmatrix}. \tag{29}$$

Therefore, when the auctioneer receives the constraint matrix $A'$, it has no way of knowing whether the agent has an MDP with transition function (29) that was transformed using (26) or the MDP with transition function (24) that was not transformed. Notice that the dynamics of the two MDPs vary significantly: both in transition probabilities and state connectivity. The second MDP does not reveal any information about the originating MDP and the corresponding market dynamics. □

To sum up, we can, to a large extent, maintain information privacy in our mechanism by allowing agents to apply linear transformations to their original LPs. The information that is revealed by our mechanism consists of agents' resource costs $\rho^m(a, o)$, capacity bounds $\widehat{\kappa}^m(c)$, and the sizes of their state and action spaces (the latter can be hidden by adding dummy states and actions to the MDP).

The revealed information can be used to infer agents' preferences and resource requirements. Further, numeric policies are revealed, but the lack of information about transition and reward functions renders this information worthless (as just illustrated in Example 8, there could be multiple originating MDPs with very different properties).

## 5. Experimental Results

In this section we present an empirical analysis of the computational complexity of the resource-allocation mechanism described in Section 4. We report results on the computational complexity of the mechanism from Section 4.2, where the agents submit their MDPs to the auctioneer, who then simultaneously solves the resource-allocation and policy-optimization problems. As far as the additional speedup achieved by distributing the WDP, as described in Section 4.3, we do not report empirical results, since it is well-established in the parallel programming literature that parallel versions of branch-and-bound MILP solvers consistently achieve linear speedup (Eckstein, Phillips, & Hart, 2000). This is due to the fact that branch-and-bound algorithms require very little inter-process communication.

For our experiments, we implemented a multiagent delivery problem, based on the multiagent rover domain (Dolgov & Durfee, 2004). In this problem, agents operate in a stochastic grid world with delivery locations randomly placed throughout the grid. Each delivery task requires a set of resources, and there are limited quantities of the resources. There are random delivery locations on the grid, and each location has a set of deliveries that it accepts. Each resource has some size requirements (capacity cost), and each delivery agent has bounded space to hold the resources (limited capacity). The agents participate in an auction where they bid on delivery resources. In this setting, the value of a resource depends on what other resources the agent acquires and what other deliveries it can make. Given a bundle of resources, an agent's policy optimization problem is to find the optimal delivery plan. The exact parameters used in our experiments are not critical for the trends seen in the results presented below, but for the sake of reproducibility the domain is described in detail in Appendix C.[13] All of the resource costs in the experiments presented below are binary.

Computational complexity of constrained optimization problems can vary greatly as constraints are tightened or relaxed. Therefore, as our first step in the analysis of empirical computational complexity of our mechanism, we investigate how its running time depends on the capacity constraints of each agent and on the bounds on the total amounts of resources shared by the agents. As is common with other types of constrained optimization or constraint-satisfaction problems, it is natural to expect that the WDP MILP will be easy to solve when the problem is over- or under-constrained in either the capacity or the resource bounds. To empirically verify this, we varied local capacity constraint levels from 0 (meaning agents cannot use any resources) to 1 (meaning each agent has the capacity to use enough resources to execute its optimal unconstrained policy), as well as the global constraint levels for which 0 meant that no resources were available to the agents, and 1 meant that there were enough resources to assign to each agent its most desired resource bundle. In all of our experiments, the part of the MILP solver was played by CPLEX 8.1 on a Pentium-4 machine with 2GB of RAM (RAM was not a bottleneck due to the use of sparse matrix representations). A typical running-time profile is shown in Figure 4. The problem is very easy when it is over-constrained, becomes more difficult as the constraints are relaxed and then abruptly becomes easy again when capacity and resource levels start to approach utopia.

---

13. We also investigated other, randomly generated domains, and the results were qualitatively the same.

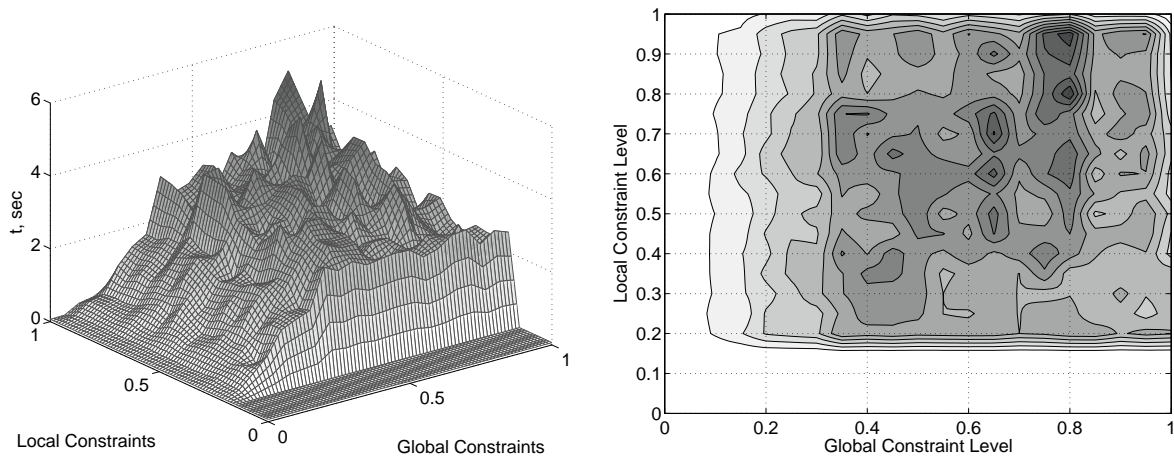

Figure 4: Running time for MDP-based winner-determination MILP (19) for different lev-
els of global $(\widehat{\rho})$ and local $(\widehat{\kappa}^m)$ constraints. The constraint levels are fractions
of utopian levels that are needed to implement optimal unconstrained policies.
Problems involved 10 agents, each operating on a 5-by-5 grid, with 10 shared
resource types. Each data point shown is an average of ten runs over randomly-
generated problems.

In all of the following experiments we aim to avoid the easy regions of constraint levels.
Therefore, given the complexity profiles, we set the constraint levels to 0.5 for both local
capacity and global resource bounds. We also set the discount factor to $\gamma = 0.95$. This
value was chosen arbitrarily, because our investigations into the effect of the value of $\gamma$ on
the running time of the MILP revealed no significant trends.

We begin by comparing the performance of our MDP-based auction (Section 4.2) to the
performance of the straightforward CA with flat preferences (as described in Section 4.1).
The results are summarized in Figure 5, which compares the time it takes to solve the
standard winner-determination problem on the space of all resource bundles (16) to the
time needed to solve the combined MDP-WDP problem (19) used in our mechanism, as
the number of resources is increased (with 5 agents, on a 5-by-5 grid). Despite the fact
that both algorithms have exponential worst-case running time, the number of integer
variables in (16) is exponentially larger than in our MILP (19), the effect of which is clearly
demonstrated in Figure 5. Furthermore, this comparison gives an extremely optimistic
view of the performance of the standard CA, as it does not take into account the additional
complexity of the valuation problem, which requires formulating and solving a very large
number of MDPs (one per resource bundle). On the other hand, the latter is embedded into
the WDP of our mechanism (19), thus including the time for solving the valuation problem
in the comparison would only magnify the effect. In fact, in our experiments, the time
required to solve the MDPs for the valuation problem was significantly greater than the
time for solving the resulting WDP MILP. However, we do not present quantitative results
to that effect here, because of the difference in implementation (iterating over resource
bundles and solving MDPs was done via a straightforward implementation in Matlab, while MILPs were solved using highly-optimized CPLEX code). No parallelization of the WDP was performed for these experiments for either algorithm.

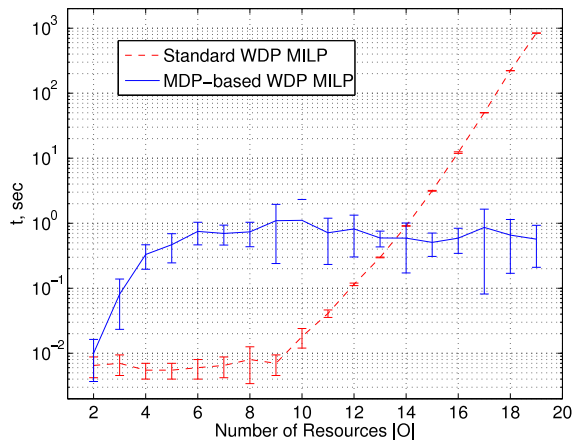

Figure 5: Gains in computation efficiency: MDP-based WDP versus a WDP in a straight-forward CA implementation. The latter does not include the time for solving the MDPs to compute resource-bundle values. Error bars show the standard deviation over ten runs.

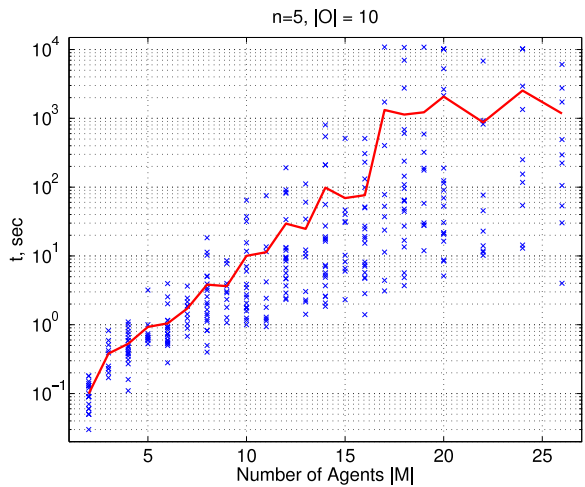

Figure 6: Scaling the MDP-based winner-determination MILP (19) to more agents. Agents operated on 5-by-5 grids and shared 10 types of resources.

Below we analyze the performance of our algorithm on larger problems infeasible for the straightforward CA. Figure 6 illustrates the scaling effect as the number of agents participating in the auction is increased. Here and below, each point on the plot corresponds to a single run of the experiment (with no less than ten runs performed for every value of parameters), and the solid line is the mean. Recall that the size of the WDP scales linearly with the number of agents. The graph therefore reflects a rather standard scaling effect for an NP-complete problem. As can be seen from the plot, problems with 25 agents and 10 resource types are well within the reach of the method, on average taking around 30 minutes.

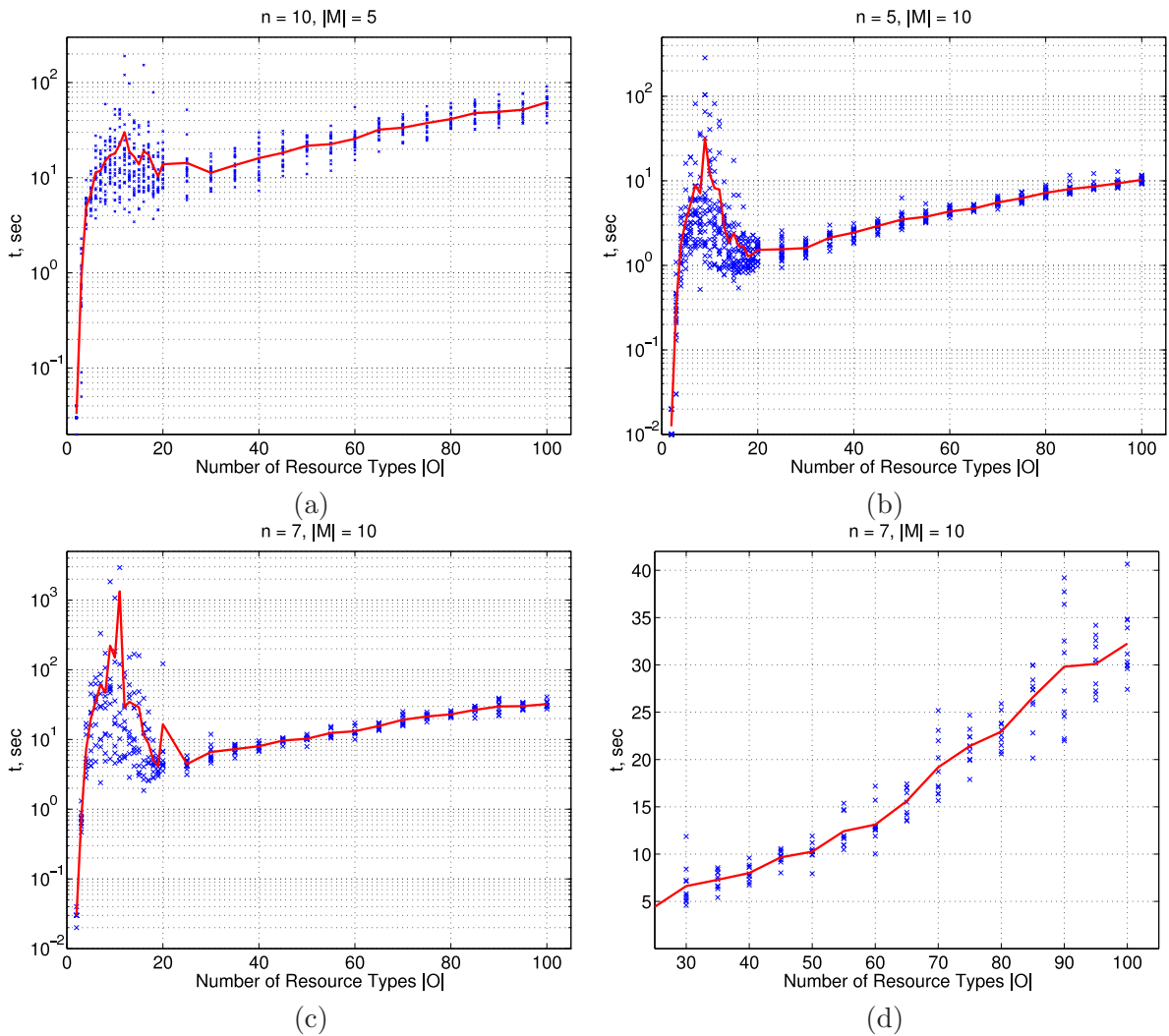

Figure 7: (a)–(c): scaling of the MDP-based winner-determination MILP (19) with the number of resources on three sets of problems with different grid sizes ($n$) and different numbers of agents ($|\mathcal{M}|$); (d): a linear-scale plot of the tail of the data in (c).

Next, we analyze how the method scales with the number of resource types. Figure 7 shows the solution time as a function of the number of resource types for three different sets of problems. In these problems, the number of actions scaled linearly with the number of resource types, but each action required a constant number of resources, i.e., the number of nonzero $\rho(a, o)$ per action was constant (two) regardless of the total number of resource types. These problems exhibit an interesting trait wherein the running time peaks for relatively low numbers of resource types, then falls quickly, and then increases much more slowly as the number of resource types increases (as illustrated in Figure 7d, which uses a linear scale). This is due to the fact that when the total number of resource types is much higher than the number of resources required by any action, there is less contention for a particular resource among the agents and between one agent's actions. Therefore, the problems become relatively under-constrained and the solution time increases slowly.

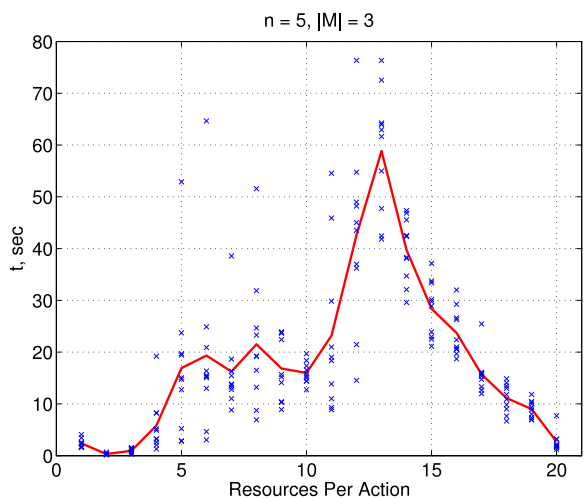

Figure 8: Complexity of MDP-based winner-determination MILP (19) as a function of the number of actions' resource requirements.

To better illustrate this effect, we ran a set of experiments inverse to the ones shown in Figure 7: we kept the total number of resource types constant and increased the number of resource types required by each action. The results are shown in Figure 8. The running-time profile is similar to what we observed earlier when we varied the local and global constraints: when the total number of resources per action is low or high, the problem is under- or over-constrained and is relatively easy to solve, but its complexity increases significantly when the number of resources required by each resource is in the range of 50-80% of the total number of resource types.

Based on the above, we would expect that if the actions' resource requirements increased with the total number of resource types, the problem would not scale as gracefully as in Figure 7. For example, Figure 9 illustrates the running time for problems where the number of resources required by each action scales linearly with the total number of resources. There, the complexity does increase significantly faster. However, it is not unreasonable to assume that in many domains the number of actions does not, in fact, increase with the total number of resource types involved. Indeed, it is natural to assume that the total number of resource types increases as the problem becomes more complicated and the number of tasks the agent can perform increases. However, why should the resource requirements of an action increase as well? If the delivery agent from our running example acquires the ability to deliver pizza, it might need new resources to perform actions related to this new activity, but one would not expect the resource requirements for delivering furniture or appliances to change. Therefore, we believe that in many real applications, our method will scale up gracefully with the total number of resource types.

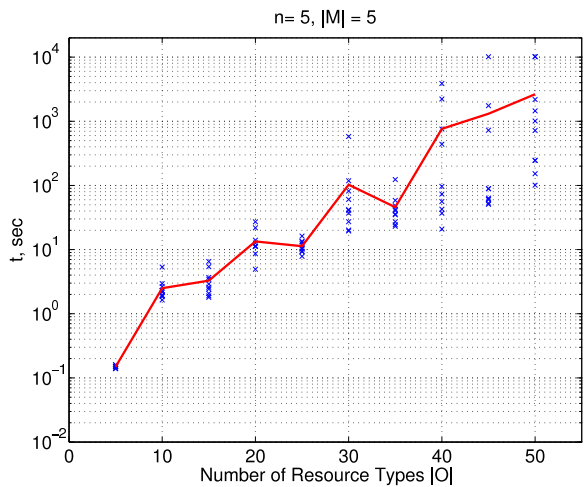

Figure 9: Complexity when actions' resource requirements grow proportionally to the total number of resource types. The number of resource types needed by each action is 10% of the total number of resource types $|\mathcal{O}|$.

The above experiments illustrate the point that for domains where agents have preferences that are defined by underlying Markov decision processes, the resource-allocation mechanism developed in this paper can lead to significant computational advantages. As shown in Figure 7, the method can be successfully applied to very large problems that, we argue, are well beyond the reach of combinatorial resource-allocation mechanisms with flat preferences. As our experiments show (Figure 5), even for small problems, combinatorial resource allocation mechanisms with flat preferences can be time-consuming, and our attempts to empirically evaluate those simpler mechanisms on larger problems proved futile. For instance, our method takes under one minute to solve a problem that, in the standard CA, requires the agents to enumerate up to $2^{100}$ bundles and the auctioneer to solve an NP-complete problem with an input of that size.

## 6. Generalizations, Extensions, and Conclusions

There are many possible extensions and generalizations of the work presented here, and we briefly outline several below.

The treatment in this paper focused on the problem of resource allocation among self-interested agents, but the algorithms also apply to cooperative MDPs with weakly-interacting agents. In the cooperative setting, the concept of resources can be viewed as a compact way to model inter-agent constraints and their inability to include some combinations of joint actions in their policies. Such weakly-coupled MDPs, where agents have independent transition and reward functions, but certain combinations of joint actions are not feasible is a widely used model of agents' interactions (e.g., Singh & Cohn, 1998). Our model was resource-centric, but more direct models are also certainly possible. For example, agents can use SAT formulas to describe valid combinations of joint actions. This case can be easily handled via simple modifications to the single and multiagent MILPs (15) and (19). Indeed, any SAT formula can be expressed as a set of linear inequalities on binary variables $\Delta(a)$ (or $\Delta^m(a)$ in the multiagent case), which can be directly added to the corresponding MILP (see the case of non-binary resources in Appendix B for an MILP defined on indicators $\Delta(a)$, instead of the $\delta(o)$ used in the binary case).

As mentioned previously, our work can be extended to handle consumable resources that are used up whenever agents execute actions. In fact, under some conditions, the problem can be considerably simplified for domains with only these kinds of resources. The most important change is that we have to redefine the value of a particular resource bundle to an agent. The difficulty is that, given a policy, the total use of consumable resources is uncertain, and the definition of the value of a resource bundle becomes ambiguous. One possibility is to define the value of a bundle as the payoff of the best policy whose *expected* resource usage does not exceed the amounts of resources in the bundle. The interpretation of $\rho^m(a, o)$ would also change to mean the amount of resource $o$ consumed by action $a$ every time it is executed. This would make the constraints in (19) *linear* in the occupation measure, which would tremendously simplify the WDP (making it polynomial). This is analogous to the models used in constrained MDPs (Altman & Shwartz, 1991), as briefly described earlier in Section 2. Information privacy can be handled similarly to the case of non-consumable resources. However, given the transformation $\mathbf{y} = D\mathbf{x}$, the resource cost function $\rho^m$ will also have to be scaled by $D^{-1}$ (since the total consumption of consumable resources is proportional to the occupation measure). This has the additional benefit of hiding the resource cost functions (unlike the case of non-consumable resources where they were revealed). A more detailed treatment of the model with consumable resources is presented in the work by Dolgov (2006), including a discussion of risk-sensitive cases, where the value of a resource bundle is defined as the payoff of the best policy whose probability of exceeding the resource amounts is bounded.

In this work we exploited structure in agents' preferences that stems from the underlying policy-optimization problems. However, the latter were modeled using "flat" MDPs that enumerate all possible states and actions. Such flat MDPs do not scale well due to the curse of dimensionality (Bellman, 1961). To address this, the WDP MILP can be modified to work with *factored* MDPs (Boutilier, Dearden, & Goldszmidt, 1995) by using a factored resource-allocation algorithm (Dolgov & Durfee, 2006), which is based on the dual ALP method for solving factored MDPs as developed by Guestrin (2003). This method allows us to exploit both types of structure in the resource-allocation algorithms: structure in agents' preferences induced by the underlying MDPs, as well as structure in MDPs themselves.

The resource-allocation mechanism discussed in this paper assumed a one-shot allocation of resources and a static population of agents. An interesting extension of our work would be to consider a system where agents and resources arrive and depart dynamically, as in the online mechanism design work (Parkes & Singh, 2003; Parkes, Singh, & Yanovsky, 2004). Combining the MDP-based model of utility functions with the dynamics of online problems could be a valuable result and thus appears to be a worthwhile direction of future work. If the agent population is static, but a periodic re-allocation of resources is allowed, techniques like *phasing* can be used to solve the resulting problem (Wu & Durfee, 2005).

To summarize the results of this paper, we presented a variant of a combinatorial auction for resource allocation among self-interested agents whose valuations of resource bundles are defined by their weakly-coupled constrained MDPs. For such problems, our mechanism, which exploits knowledge of the structure of agents' MDP-based preferences, achieves an exponential reduction in the number of integer decision variables, which in turn leads to tremendous speedup over a straightforward implementation, as confirmed by our experimental results. Our mechanism can be implemented to achieve its reduction in computational complexity without sacrificing any of the nice properties of a VCG mechanism (optimal outcomes, strategy-proofness, and voluntary participation). We also discussed a distributed implementation of the mechanism that retains strategy-proofness (using the fact that an LP solution can be easily verified), and does not reveal agents' private MDP information (using a transformation of agents' MDPs).

We believe that the models and solution algorithms described in this paper significantly further the applicability of combinatorial resource-allocation mechanisms to practical problems, where the utility functions for resource bundles are defined by sequential stochastic decision-making problems.

## 7. Acknowledgments

We thank the anonymous reviewers for their helpful comments, as well as our colleagues Satinder Singh, Kang Shin, Michael Wellman, Demothenis Teneketsis, Jianhui Wu, and Jeffrey Cox for the valuable discussions related to this work.

This material is based in part upon work supported by Honeywell International, and by the DARPA IPTO COORDINATORs program and the Air Force Research Laboratory under Contract No. FA8750–05–C–0030. The views and conclusions contained in this document are those of the authors, and should not be interpreted as representing the official policies, either expressed or implied, of the Defense Advanced Research Projects Agency or the U.S. Government.

## Appendix A. Proofs

### A.1 Proof of Theorem 1

**Theorem 1** *Consider a finite set of $n$ indivisible resources $\mathcal{O} = \{o_i\}$ ($i \in [1, n]$), with $m \in \mathbb{N}$ available units of each resource. Then, for any non-decreasing utility function defined over resource bundles $f : [0, m]^n \mapsto \mathbb{R}$, there exists a resource-constrained MDP $\langle \mathcal{S}, \mathcal{A}, p, r, \mathcal{O}, \rho, \mathcal{C}, \kappa, \widehat{\kappa}, \alpha \rangle$ (with the same resource set $\mathcal{O}$) whose induced utility function over the resource bundles is the same as $f$. In other words, for every resource bundle $\mathbf{z} \in [0, m]^n$, the value of the optimal policy among those whose resource requirements do not exceed $\mathbf{z}$ (call this set $\Pi(\mathbf{z})$) is the same as $f(\mathbf{z})$:*

$$\forall f q : [0, m]^n \mapsto \mathbb{R}, \ \exists \ \langle \mathcal{S}, \mathcal{A}, p, r, \mathcal{O}, \rho, \mathcal{C}, \kappa, \widehat{\kappa}, \alpha \rangle :$$

$$\forall \ \mathbf{z} \in [0, m]^n, \Pi(\mathbf{z}) = \left\{ \pi \Big| \max_a \Big[ \rho(a, o_i) H \big( \sum_s \pi(s, a) \big) \Big] \leq z_i \right\} \implies \max_{\pi \in \Pi(\mathbf{z})} U_\gamma(\pi, \alpha) = f(\mathbf{z}).$$

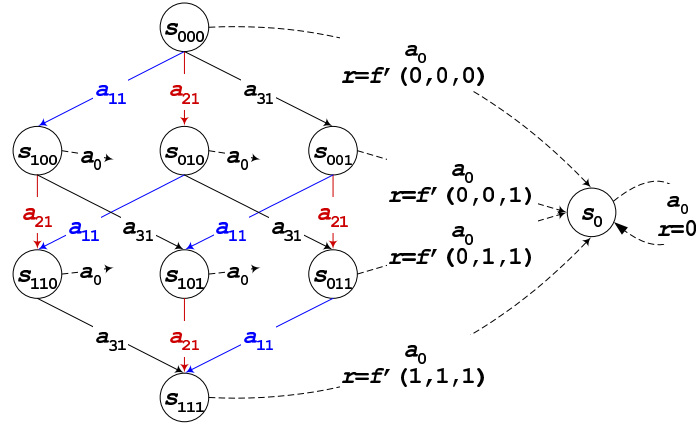

Figure 10: Creating an MDP with resources for an arbitrary non-decreasing utility function. The case shown has three binary resources. All transitions are deterministic.

**Proof:** This statement can be shown via a straightforward construction of an MDP that has an exponential number (one per resource bundle) of states or actions. Below we present a reduction with a linear number of actions and an exponential number of states. Our choice is due to the fact that, although the reverse mapping requiring two states and exponentially many actions is even more straightforward, such an MDP feels somewhat unnatural.

Given an arbitrary non-decreasing utility function $f$, a corresponding MDP can be constructed as follows (illustrated in Figure 10 for $n = 3$ and $m = 1$). The state space $\mathcal{S}$ of the MDP consists of $(m+1)^n + 1$ states – one state ($s_{\mathbf{z}}$) for every resource bundle $\mathbf{z} \in [0, m]^n$, plus a sink state ($s_0$).

The action space of the MDP $\mathcal{A} = a_0 \bigcup \{a_{ij}\}$, $i \in [1, n]$, $j \in [1, m]$ consists of $mn + 1$ actions: $m$ actions per each resource $o_i$, $i \in [1, n]$, plus an additional action $a_0$.

The transition function $p$ is deterministic and is defined as follows. Action $a_0$ is applicable in every state and leads to the sink state $s_0$. Every other action $a_{ij}$ is applicable in states $s_{\mathbf{z}}$, where $z_i = (j - 1)$ and leads with certainty to the states where $z_i = j$:

$$p(\sigma|s, a) = \begin{cases} 1 & a = a_{ij}, s = s_{\mathbf{z}}, \sigma = s_{\mathbf{z}'}, z_i = (j-1), z_i' = j, \\ 1 & a = a_0, \sigma = s_0, \\ 0 & \text{otherwise.} \end{cases}$$

In other words, $a_{ij}$ only applies in states that have $j - 1$ units of resource $i$ and leads to the state where the amount of $i^{\text{th}}$ resource increases to $j$.

The reward function $r$ is defined as follows. There are no rewards in state $s_0$, and action $a_0$ is the only action that produces rewards in other states:

$$r(s, a) = \begin{cases} f'(\mathbf{z}) & a = a_o, s = s_{\mathbf{z}}, \ \forall \mathbf{z} \in [0, m]^n \\ 0 & \text{otherwise,} \end{cases}$$

where $f'$ is a simple transformation of $f$ that compensates for the effects of discounting:

$$f'(\mathbf{z}) = f(\mathbf{z})(\gamma)^{-\sum_i z_i}.$$

In other words, it takes $\sum_i z_i$ transitions to get to state $s_{\mathbf{z}}$, so the contribution of the above into the total discounted reward will be exactly $f(\mathbf{z})$.

The resource requirements $\rho$ of actions are as follows: action $a_0$ does not require any resources, while every other action $a_{ij}$ requires $j$ units of resource $o_i$.

Finally, the initial conditions are $\alpha(s_{\mathbf{z}=\mathbf{0}}) = 1$, meaning that the agent always starts in the state that corresponds to the empty resource bundle (state $s_{000}$ in Figure 10). The capacity costs $\kappa$ and limits $\widehat{\kappa}$ are not used, so we set $\mathcal{C} = \varnothing$.

It is easy to see that in the MDP constructed above, given a resource bundle $\mathbf{z}$, any policy from the feasible set $\Pi(\mathbf{z})$ has zero probability of reaching any state $s_{\mathbf{z}'}$ for which $\mathbf{z}' > \mathbf{z}$ (for any component $i$). Furthermore, an optimal policy from the set $\Pi(\mathbf{z})$ will be to transition to state $s_{\mathbf{z}}$ (since $f(\mathbf{z})$ is non-decreasing) and then use action $a_0$, thus obtaining a total discounted reward of $f(\mathbf{z})$. $\qquad\square$

## A.2 Proof of Theorem 2

**Theorem 2** *Given an MDP $M = \langle \mathcal{S}, \mathcal{A}, p, r, \mathcal{O}, \rho, \mathcal{C}, \kappa, \widehat{\kappa}, \alpha \rangle$ with resource and capacity constraints, if there exists a policy $\pi \in \Pi^{\mathrm{HR}}$ that is a feasible solution for $M$, there exists a stationary deterministic policy $\pi^{\mathrm{SD}} \in \Pi^{\mathrm{SD}}$ that is also feasible, and the expected total reward of $\pi^{\mathrm{SD}}$ is no less than that of $\pi$:*

$$\forall \, \pi \in \Pi^{\mathrm{HR}}, \; \exists \, \pi^{\mathrm{SD}} \in \Pi^{\mathrm{SD}} : U_\gamma(\pi^{\mathrm{SD}}, \alpha) \geq U_\gamma(\pi, \alpha)$$

**Proof:** Let us label $\mathcal{A}' \subseteq \mathcal{A}$ the set of all actions that have a non-zero probability of being executed according to $\pi$, i.e.,

$$\mathcal{A}' = \{a | \exists s : \pi(s, a) > 0\}$$

Let us also construct an unconstrained MDP: $M' = \langle \mathcal{S}, \mathcal{A}', p', r' \rangle$, where $p'$ and $r'$ are the restricted versions of $p$ and $r$ with the action domain limited to $\mathcal{A}'$:

$$p' : \mathcal{S} \times \mathcal{A}' \times \mathcal{S} \mapsto [0, 1]$$
$$r' : \mathcal{S} \times \mathcal{A}' \mapsto \mathbb{R}$$
$$p'(\sigma|s, a) = p(\sigma|s, a), \; r'(s, a) = r(s, a) \quad \forall s \in \mathcal{S}, \sigma \in \mathcal{S}, a \in \mathcal{A}'$$

Due to a well-known property of unconstrained infinite-horizon MDPs with the total expected discounted reward optimization criterion, $M'$ is guaranteed to have an optimal stationary deterministic solution (e.g., Theorem 6.2.10, Puterman, 1994), which we label $\pi^{\mathrm{SD}}$.

Consider $\pi^{\mathrm{SD}}$ as a potential solution to $M$. Clearly, $\pi^{\mathrm{SD}}$ is a feasible solution, because its actions come from the set $\mathcal{A}'$ that includes actions that $\pi$ uses with non-zero probability, which means that the resource requirements (as in (9)) of $\pi^{\mathrm{SD}}$ can be no greater than those of $\pi$. Indeed:

$$\max_{a \in \mathcal{A}'} \left\{ \rho(a, o) H\left(\sum_s \pi^{\mathrm{SD}}(s, a)\right) \right\} \leq \max_{a \in \mathcal{A}'} \rho(a, o) = \max_{a \in \mathcal{A}} \left\{ \rho(a, o) H\left(\sum_s \pi(s, a)\right) \right\}, \qquad (30)$$

where the first inequality is due to the fact that $H(z) \leq 1 \; \forall z$, and the second equality follows from the definition of $\mathcal{A}'$.

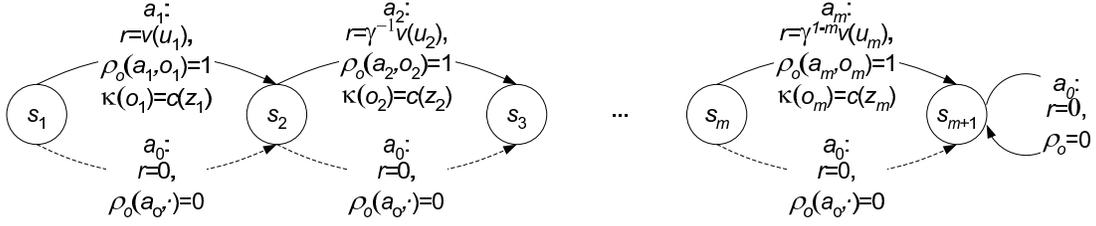

Figure 11: Reduction of KNAPSACK to M-OPER-CMDP. All transitions are deterministic.

Furthermore, observe that $\pi^{\text{SD}}$ yields the same total reward under $M'$ and $M$. Additionally, since $\pi^{\text{SD}}$ is a uniformly optimal solution to $M'$, it is, in particular, optimal for the initial conditions $\alpha$ of the constrained MDP $M$. Therefore, $\pi^{\text{SD}}$ constitutes a feasible solution to $M$ whose expected reward is greater than or equal to the expected reward of any feasible policy $\pi$. □

### A.3 Proof of Theorem 3

**Theorem 3** *The following decision problem is NP-complete. Given an instance of an MDP $\langle \mathcal{S}, \mathcal{A}, p, r, \mathcal{O}, \rho, \mathcal{C}, \kappa, \widehat{\kappa}, \alpha \rangle$ with resources and capacity constraints, and a rational number $Y$, does there exist a feasible policy $\pi$, whose expected total reward, given $\alpha$, is no less than $Y$?*

**Proof:** As shown in Theorem 2, there always exists an optimal policy for (9) that is stationary and deterministic. Therefore, the presence in NP is obvious, since we can, in polynomial time, guess a stationary deterministic policy, verify that it satisfies the resource constraints, and calculate its expected total reward (the latter can be done by solving the standard system of linear Markov equations (2) on the values of all states).

To show NP-completeness of the problem, we use a reduction from KNAPSACK (Garey & Johnson, 1979). Recall that KNAPSACK in an NP-complete problem, which asks whether, for a given set of items $z \in \mathcal{Z}$, each of which has a cost $c(z)$ and a value $v(z)$, there exists a subset $\mathcal{Z}' \subseteq \mathcal{Z}$ such that the total value of all items in $\mathcal{Z}'$ is no less than some constant $\widehat{v}$, and the total cost of the items is no greater than another constant $\widehat{c}$, i.e., $\sum_{z \in \mathcal{Z}'} c(z) \leq \widehat{c}$ and $\sum_{z \in \mathcal{Z}'} v(z) \geq \widehat{v}$. Our reduction is illustrated in Figure 11 and proceeds as follows.

Given an instance of KNAPSACK with $|\mathcal{Z}| = m$, let us number all items as $z_i$, $i \in [1, m]$ as a notational convenience. For such an instance of KNAPSACK, we create an MDP with $m + 1$ states $\{s_1, s_2, \ldots s_{m+1}\}$, $m + 1$ actions $\{a_0, \ldots a_m\}$, $m$ types of resources $\mathcal{O} = \{o_1, \ldots o_m\}$, and a single capacity $\mathcal{C} = \{c_1\}$.

The transition function on these states is defined as follows. Every state $s_i$, $i \in [1, m]$ has two transitions from it, corresponding to actions $a_i$ and $a_0$. Both actions lead to state $s_{i+1}$ with probability 1. State $s_{m+1}$ is absorbing and all transitions from it lead back to itself.

The reward and the cost functions are defined as follows. We want action $a_i$, $i \in [1, m]$ (which corresponds to item $z_i$ in KNAPSACK) to contribute $v(z_i)$ to the total discounted reward. Hence, we set the immediate reward for every action $a_i$ to $v(z_i)(\gamma)^{1-i}$, which, given that our transition function implies that state $s_i$ is reached exactly at step $i - 1$, ensures that if action $a_i$ is ever executed, its contribution to the total discounted reward will be $v(z_i)(\gamma)^{1-i}(\gamma)^{i-1} = v(z_i)$. Action $a_0$ produces a reward of zero in all states.

The resource requirements of actions are defined as follows. Action $a_i$, $i \in [1, m]$ only needs resource $o_i$, i.e., $\rho(a_i, o_j) = 1 \iff i = j$. We set the cost of resource $o_i$ to be the cost $c(z_i)$ of item $i$ in the KNAPSACK problem. The "null" action $a_0$ requires no resources.

In order to complete the construction, we set the initial distribution $\alpha = [1, 0, \ldots]$ so that the agent starts in state $s_1$ with probability 1. We also define the decision parameter $Y = \widehat{v}$ and the upper bound on the single capacity $\widehat{\kappa} = \widehat{c}$.

Essentially, this construction allows the agent to choose action $a_i$ or $a_0$ at every state $s_i$. Choosing action $a_i$ is equivalent to putting item $z_i$ into the knapsack, while action $a_0$ corresponds to the choice of not including $z_i$ in the knapsack. Therefore, there exists a policy that has the expected payoff no less than $Y = \widehat{v}$ and uses no more than $\widehat{\kappa} = \widehat{c}$ of the shared resource if and only if there exists a solution to the original instance of KNAPSACK. □

## Appendix B. Non-binary Resource Requirements

Below we describe an MILP formulation of the capacity-constrained single-agent optimization problem (9) for arbitrary resource costs $\rho : \mathcal{A} \times \mathcal{O} \mapsto \mathbb{R}$, as opposed to binary costs that were assumed in the main parts of the paper. The corresponding multiagent winner-determination problem (the non-binary equivalent of (19)) follows immediately from the single-agent MILP.

For arbitrary resource costs, we obtain the following non-binary equivalent of the optimization problem (12) in the occupation measure coordinates:

$$\max_x \sum_s \sum_a x(s, a) r(s, a)$$

subject to:

$$\sum_a x(\sigma, a) - \gamma \sum_s \sum_a x(s, a) p(\sigma|s, a) = \alpha(\sigma), \qquad \forall \sigma \in \mathcal{S};$$

$$\sum_o \kappa(o, c) \max_a \left\{ \rho(a, o) H\left(\sum_s x(s, a)\right) \right\} \leq \widehat{\kappa}(c), \qquad \forall c \in \mathcal{C}; \qquad (31)$$

$$x(s, a) \geq 0, \qquad \forall s \in \mathcal{S}, a \in \mathcal{A}.$$

To linearize the sum of max operators in (31), let us observe that the inequality

$$\sum_i^n g(u_i) \max_{z \in \mathcal{Z}} f(z, u_i) = g(u_1) \max_{z \in \mathcal{Z}} f(z, u_1) + \ldots + g(u_n) \max_{z \in \mathcal{Z}} f(z, u_n) \leq a$$

is equivalent to the following system of $|\mathcal{Z}|^n$ linear inequalities:

$$g(u_1) f(z_1, u_1) + g(u_2) f(z_2, u_2) + \ldots + g(u_n) f(z_n, u_n) \leq a, \qquad \forall z_1, z_2, \ldots z_n \in \mathcal{Z}.$$

Applying this to the constraints from (31), we can express the original system of $|\mathcal{C}|$ nonlinear constraints (each of which has a max):

$$\sum_o \kappa(o,c) \max_a \left\{ \rho(a,o) H\big( \sum_s x(s,a) \big) \right\} \leq \widehat{\kappa}(c), \qquad \forall c \in \mathcal{C}$$

as the following system of $|\mathcal{C}||\mathcal{A}|^{|\mathcal{O}|}$ constraints where the max is removed:

$$\sum_o \kappa(o,c) \rho(a_o, o) H\big( \sum_s x(s,a) \big) \leq \widehat{\kappa}(c), \qquad \forall c \in \mathcal{C}, a_{o_1}, a_{o_2}, \ldots \in \mathcal{A}. \qquad (32)$$

Notice that this way of eliminating the maximization exponentially increases the number of constraints, because the above expansion enumerates all possible actions for each resource (i.e., it enumerates policies where each resource $o$ is used by action $a_1$, where it is used by action $a_2$, action $a_3$, etc.) However, in many problems not all resources are used by all actions. In such cases, most of the above constraints become redundant, and the number of constraints can be reduced from $|\mathcal{C}||\mathcal{A}|^{|\mathcal{O}|}$ to $|\mathcal{C}| \prod_o |\mathcal{A}_o|$, where $\mathcal{A}_o$ is the number of actions that use resource $o$.

We can linearize the Heaviside function analogously to the case of binary resource costs in Section 3.2: we create a binary indicator variable that corresponds to the argument of $H()$ and tie it to the occupation measure $x$ via linear inequalities. The only difference is that for non-binary resource costs, instead of using indicators on resources, we use indicators on actions: $\Delta(a) \in \{0,1\}$, where $\Delta(a) = H(\sum_s x(s,a))$ is an indicator that shows whether action $a$ is used in the policy. Using $\Delta$ and expanding the max as above, we can represent the optimization problem (9) as the following MILP:

$$\max_{x,\Delta} \sum_s \sum_a x(s,a) r(s,a)$$

subject to:

$$\sum_a x(\sigma,a) - \gamma \sum_s \sum_a x(s,a) p(\sigma|s,a) = \alpha(\sigma), \qquad \forall \sigma \in \mathcal{S};$$

$$\sum_o \kappa(o,c) \rho(a_o, o) \Delta(a_o) \leq \widehat{\kappa}(c), \qquad \forall c \in \mathcal{C}, a_{o_1}, a_{o_2}, \ldots \in \mathcal{A}; \qquad (33)$$

$$\sum_s x(s,a)/X \leq \Delta(a), \qquad \forall a \in \mathcal{A};$$

$$x(s,a) \geq 0, \qquad \forall s \in \mathcal{S}, a \in \mathcal{A};$$

$$\Delta(a) \in \{0,1\}, \qquad \forall a \in \mathcal{A},$$

where $X \geq \max \sum_s x(s,a)$ is some constant finite upper bound on the expected number of times action $a$ is used, which exists for any discounted MDP. We can, for example, let $X = (1-\gamma)^{-1}$, since $\sum_{s,a} x(s,a) = (1-\gamma)^{-1}$ for any $x$ that is a valid occupation measure for an MDP with discount factor $\gamma$.

**Example 9** *Let us formulate the MILP for the constrained problem from Example 3. Recall that there are three resources $\mathcal{O} = \{o_t, o_f, o_m\}$ (truck, forklift, and mechanic), one capacity* type $\mathcal{C} = \{c_1\}$ (money), and actions have the following resource requirements (listing only the nonzero ones):

$$\rho(a_1, o_t) = 1, \ \rho(a_2, o_t) = 1, \ \rho(a_2, o_f) = 1, \ \rho(a_3, o_t) = 1, \ \rho(a_4, o_t) = 1, \ \rho(a_4, o_m) = 1$$

The resources have the following capacity costs:

$$\kappa(o_t, c_1) = 2, \ \kappa(o_f, c_1) = 3, \ \kappa(o_m, c_1) = 4,$$

and the agent has a limited budget, i.e., a capacity bound $\widehat{\kappa}(c_1) = 8$.

To compute the optimal policy for an arbitrary $\alpha$, we can formulate the problem as an MILP using the techniques described above. Using binary variables $\{\Delta(a_i)\} = \{\Delta_i\} = \{\Delta_1, \Delta_2, \Delta_3, \Delta_4\}$,[14] we can express the constraint on capacity cost as the following system of $|\mathcal{C}||\mathcal{A}|^{|\mathcal{O}|} = 1(4)^3 = 64$ linear constraints:

$$(2)(1)\Delta_1 + (3)(0)\Delta_1 + (4)(0)\Delta_1 \le 8,$$
$$(2)(1)\Delta_1 + (3)(0)\Delta_1 + (4)(0)\Delta_2 \le 8,$$
$$(2)(1)\Delta_1 + (3)(0)\Delta_1 + (4)(0)\Delta_3 \le 8,$$
$$(2)(1)\Delta_1 + (3)(0)\Delta_1 + (4)(1)\Delta_4 \le 8,$$
$$(2)(1)\Delta_1 + (3)(1)\Delta_2 + (4)(0)\Delta_1 \le 8,$$
$$\dots$$
$$(2)(0)\Delta_4 + (3)(0)\Delta_4 + (4)(1)\Delta_4 \le 8.$$

It is easy to see that most of these constraints are redundant, and the fact that each action only requires a small subset of the resources allows us to prune many of the constraints. In fact, the only resource that is used by multiple actions is $o_t$. Therefore, in accordance with our earlier discussion, we only need $\prod_o |\mathcal{A}_o| = 1 \times 4 \times 1 = 4$ constraints:

$$(2)(1)\Delta_1 + (3)(1)\Delta_2 + (4)(1)\Delta_4 \le 8,$$
$$(2)(1)\Delta_2 + (3)(1)\Delta_2 + (4)(1)\Delta_4 \le 8,$$
$$(2)(1)\Delta_3 + (3)(1)\Delta_2 + (4)(1)\Delta_4 \le 8,$$
$$(2)(1)\Delta_4 + (3)(1)\Delta_2 + (4)(1)\Delta_4 \le 8,$$

where each of the four constraints corresponds to a case where the first resource ($o_t$) is used by a different action.

As mentioned earlier, we can set $X = (1-\gamma)^{-1}$ for the constraints that synchronize the occupation measure $x$ and the binary indicators $\Delta$. Combining this with other constraints from (33), we get an MILP with 12 continuous and 4 binary variables, and $|\mathcal{S}| + |\mathcal{C}| \prod_o |\mathcal{A}_o| + |\mathcal{A}| = 3 + 4 + 3 = 10$ constraints (not counting the last two sets of range constraints). $\quad\square$

Finally, let us observe that by expanding the resource and action sets, any problem can be represented using binary resources only. If the domain contains mostly binary requirements, it may be more effective to expand the non-binary resource requirements $\rho$ by augmenting the resource set $\mathcal{O}$, and then use the binary formulation of Section 3.2 rather than directly applying the more-general formulation described above.

---

14. We do not create a $\Delta_0$ for the noop action $a_0$, as its resource costs are zero, and it drops out of all expressions.

## Appendix C. Experimental Setup

This appendix details how our experimental domains were constructed. For a delivery domain with $|\mathcal{M}|$ agents operating on an $n$-by-$n$ grid and sharing $|\mathcal{O}|$ resource types, we used the following parameters.

The resources enable agents to carry out delivery tasks. For a problem with $|\mathcal{O}|$ resource types, there are $|\mathcal{O}|$ delivery actions, and performing action $i \in [1, |\mathcal{O}|]$ requires a random subset of resources from $\mathcal{O}$ (where the number of resources required by an action is an important parameter, whose effect on complexity is discussed in Section 5). The probability that task $i \in [1, |\mathcal{O}|]$ can be carried at a location is $0.1 + 0.4(|\mathcal{O}| - i)/(|\mathcal{O}| - 1)$, i.e., uniformly distributed between 0.1 and 0.5, as a function of the action ID (actions with lower IDs are more rewarding, per the definition of the reward function below, but can be executed at fewer locations).

There are $n^2/5$ possible delivery locations randomly placed on the grid. Each delivery location is assigned a set of delivery tasks that can be executed there (a single location can be used for multiple delivery tasks, and a single task can be carried out at any of several locations). The assignment of tasks to locations is done randomly.

Each agent has $4 + |\mathcal{O}|$ actions: drive in any of the four perpendicular directions and execute one of the delivery tasks. The drive actions result in movement in the intended direction with probability of 0.8 and with probability of 0.2 produce no change of location. All movement actions incur a negative reward, the amount of which depends on the size of the agent. For a problem with $|\mathcal{M}|$ agents, the movement penalty incurred by agent $m \in [1, |\mathcal{M}|]$ is $-1 - 9(m-1)/(|\mathcal{M}| - 1)$, i.e., distributed uniformly on $[-1, -10]$ as a function of the agent's ID.

Execution of an action corresponding to a delivery task $i \in [1, |\mathcal{O}|]$ in a location to which the task is assigned produces a reward of $100i/|\mathcal{O}|$ and moves the agent to a new random location on the grid. The new location is chosen randomly at problem generation (thus known to agent), but the transition is deterministic, which induces a topology with nearby and remote locations. Attempting execution of a delivery task in an incorrect location does not change state and produces zero reward.

The agents bid for delivery resources of $|\mathcal{O}|$ types. There are $c_{glob}|\mathcal{M}|$ units of each resource, where $c_{glob}$ is the global constraint level (set to 0.5 for most of our experiments, as described in more detail in Section 5). There is one capacity type: size. The size requirements for making deliveries of type $i \in [1, |\mathcal{O}|]$ is $i$. The capacity limit of agent $m$ is $c_{loc}/2|\mathcal{O}|(|\mathcal{O}| + 1)$, where $c_{loc}$ is the local constraint level (set to 0.5 for most of our experiments, as was described in more detail in Section 5).

The initial location of each agent is randomly selected from a uniform distribution. The discount factor is $\gamma = 0.95$.